\documentclass[a4paper,11pt]{article}
\usepackage{jheppub} 
\usepackage{lineno}
\usepackage{xcolor}
\usepackage{braket}
\usepackage{slashed}
\newcommand{\beq}{\begin{equation}}
\newcommand{\eeq}{\end{equation}}
\newcommand{\bea}{\begin{eqnarray}}
\newcommand{\eea}{\end{eqnarray}}
\newcommand{\mec}

\title{\boldmath Tetraquarks at large $M$  and large $N$}

\author[a]{Héloïse Allaman,}
\emailAdd{heloise@allaman.net}
\author[a]{Majid Ekhterachian,}
\emailAdd{majid.ekhterachian@epfl.ch}
\author[a]{Filippo Nardi,}
\emailAdd{filippo.nardi@epfl.ch}
\author[a,b,c]{Riccardo Rattazzi,}
\emailAdd{riccardo.rattazzi@epfl.ch}
\author[a]{Stefan Stelzl}
\emailAdd{stefan.stelzl@epfl.ch}

\affiliation[a]{Theoretical Particle Physics Laboratory (LPTP),
Institute of Physics, EPFL, Lausanne, Switzerland}

\affiliation[b]{Center for Cosmology and Particle Physics, New York University,
New York,  
U.S.A.}
\affiliation[c]{ CERN, Theoretical Physics Department, Geneva, Switzerland}

\abstract{ 
We study tetraquarks in large $N$ QCD with heavy quarks, in the domain where non-relativistic quantum mechanics offers an adequate approximation.
Within the regime of validity of  the Born-Oppenheimer approximation, we systematically study and explicitly construct  tetraquark states.
At leading order in the $1/N$ expansion, the bound spectrum consists of free mesons, while the $1/N$ corrections give rise to a Born-Oppenheimer potential that can bind the mesons into tetraquarks. 
We find  two different types of tetraquarks, each endowed with distinct color-spatial wavefunctions. 
These states arise in the presence of an $\mathcal{O}(N)$ mass hierarchy between the quarks and the antiquarks. We provide a quantitative argument indicating that only for such a hierarchy is the ground state of the system a tetraquark.
We discuss what the extrapolation of our results to realistic values of the parameters may imply for the QCD tetraquark states.
}

\begin{document}
\maketitle
\flushbottom

\section{Introduction}
\label{sec:intro}

Experimental evidence for the existence of  exotic hadrons containing four or five valence quarks has been accumulating since the observation of $X(3872)$ \cite{Belle:2003nnu}. 
The  early candidate four quark states contained a heavy quark anti-quark pair (for reviews, see e.g. \cite{Esposito:2016noz,Guo:2017jvc,Olsen:2017bmm,Brambilla:2019esw}). More recently, however, states with a $cc\bar{u} \bar{d}$ flavor content, labelled $T_{cc}^+$ \cite{LHCb:2021vvq,LHCb:2021auc}, as well as a candidate $c c \bar{c}\bar{c}$ state called $X(6900)$ \cite{LHCb:2020bwg} have also been identified by the LHCb experiment. The nature of all the tetraquark states is still debated.
The future experimental program will provide more data on the existing states and may potentially lead to observation of new states such as the analogs of $T_{cc}$ involving one or two $b$ quarks or even states with more than two heavy quarks. 

The observed candidate states exhibit the peculiar feature of extreme closeness to the corresponding two meson thresholds. In particular the mass of the $X(3872)$ is within $120$ keV of the $D^0 \bar{D}^{*0}$ threshold, and  $T_{cc}^+$ has been observed to have a mass within only around $400$ keV of the $D^0 D^{*+}$ threshold.  Other candidates tetraquarks (e.g. $Z_c(3900),\, Z_c(4020),$ and $ Z_b(10650)$) are also  found within $\sim 10 $ MeV of the corresponding two meson threshold. 
Interpreting these states as compact tetraquarks, one would  expect their binding energy to be $O(\Lambda_\text{QCD})$, while for loosely bound molecular states one would expect it to roughly scale as $\Lambda_\text{QCD}^2/M$, with $M$ is the mass of the constituent mesons. Nevertheless, in both cases, the extreme closeness to threshold seems to require parametric tuning, tough at different levels.

The ongoing debate on the true nature of  tetraquarks, as well as the potential for future experimental progress, strongly motivates studying them within a controlled theoretical framework.
It is possible that, in view of the strongly coupled nature of the relevant dynamics, a full clarification will only eventually come with sufficiently accurate lattice QCD simulations. Nonetheless systematic Effective Field Theory (EFT) approaches, like Heavy Quark EFT (see e.g.~\cite{Manohar_Wise_2000}) or Non Relativistic EFT (see e.g.~\cite{Kaplan:2005es}), will surely always play a central role in the description of both the spectrum and of the phenomenology,
illustrations of the former and latter approaches can be found for instance in \cite{Eichten:2017ffp,Braaten:2020nwp}
and \cite{Fleming:2007rp,Braaten:2003he,Braaten:2015tga} respectively. Alongside these systematic approaches the large $N$ limit \cite{tHooft:1973alw} has  since long offered a qualitative or semi-quantitative, yet deep, understanding of the strong interactions. The large $N$ limit is normally considered in the strongly coupled regime, involving massless quarks and gluons. In this paper we will instead use it in conjunction with the large quark mass limit. The conjunction of the two limits will allow us to treat analytically,  within non-relativistic quantum mechanics, the non-trivial four body bound state problem. That in our mind compensates for the fact that the system we are considering is not fully realistic.
Nonetheless lessons for the real world are not excluded.

The existence of narrow tetraquarks  in the large $N$ limit of QCD  has been under debate during the last decade. Arguments for the absence of such states were originally given by Witten \cite{Witten:1979kh} and by Coleman in \cite{Coleman:1985rnk} in their classic papers on the large $N$ expansion. The main argument consists in  the observation that in the leading large $N$ approximation the two point function of tetraquark operators factorizes into the disconnected product of meson propagators, so that tetraquark poles are not found in such correlators.
That mesons (and baryons) represent the only resonances at infinite $N$ also intuitively matches the fact that mesons  are free in that limit, and thus cannot bind into tetraquarks. However in 2013
Weinberg  \cite{Weinberg:2013cfa}  pointed out a potential loophole in the main argument:  the connected part of the tetraquark 2-point function, even if subleading in the $1/N$ expansion, may still contain a tetraquark pole. Application of the LSZ approach would then allow to construct the scattering amplitudes involving this state. 
The main issue in that respect is whether its width is self-consistently suppressed at large $N$. Indeed an unsuppressed width 
would offer an additional argument against the existence of tetraquarks.
Now, as shown by Weinberg and subsenquently analyzed in more detail, the $N$ power counting of the 3-point function for  one tetraquark and two mesonic  operators
shows that the width would, self-consistently, be $1/N$ suppressed. To be more precise, arguments have also been provided indicating that a tetraquark singularity can only exist in diagrams with non-planar topology (\cite{Knecht:2013yqa,Maiani:2018pef,Cohen:2014tga}, for reviews see e.g. \cite{Ali_Maiani_Polosa_2019} or \cite{Lucha:2021mwx}). But even in that case the $N$ power counting  is consistent with a suppressed width, albeit by a different power of $1/N$.
Of course while Weinberg's remark, and the works that followed, points to a possible loophole in the arguments againts tetraquark, it unfortunately cannot provide a solid argument in favor. Our study is partly motivated by this frustrating state of things. We focus on an admittedly  more special situation with the goal to be rewarded with some solid conclusions.

There are different ways to define the large $N$ limit such that the exotic hadrons reduce  for $N=3$ to QCD tetraquarks.  
In this paper, tetraquarks are bound states involving two quarks in the fundamental representation of $SU(N)$ and two anti-quarks in the anti-fundamental.
Another  approach is to consider large $N$ QCD with 
quarks in two-index-antisymmetric representation \cite{Armoni:2003fb}, with   tetraquarks  made up of two quarks and two antiquarks \cite{Cohen:2014via}.
Yet another option is to stick to quarks in the fundamental and consider  baryonium states, made up of of $N-1$ quarks and $N-1$ antiquarks  as considered already by Witten \cite{Witten:1979kh}.

We will work in the regime where all quark masses are much above the QCD scale, treating the 't Hooft coupling $\alpha_s N $ as fixed but much smaller than $O(1)$. This will allow us to study tetraquarks benefitting from both the non-relativistic approximation and the $1/N$ expansion.
Allowing additionally for a hierarchy among the quark masses will allow us to also employ a controlled Born-Oppenheimer (BO) approximation in the  study of bound states. 
We will find that stable $QQ\bar{q}\bar{q}$ tetraquark states with two heavier quarks and two lighter antiquarks can be systematically constructed if the mass hierarchy is larger than $O(N)$. This condition can be understood as follows. At leading order in $1/N$, free mesons are the exact eigenstates of the Hamiltonian while a BO  potential for the heavier quarks  only arises as a  subleading $1/N$ correction. The latter 
 can bind the mesons only if their kinetic energy is similarly suppressed, i.e. if their mass is sufficiently large. In the regime of validity of the BO approximation the ground state of the four quark system is indeed a stable tetraquark. However our construction also entails excited states that are expected to decay mostly into mesons when considering either corrections to the BO approximantion or gluon emission. While most our explicit results pertain the BO regime, in a final section we provide evidence that no exactly stable tetraquark exists outside this regime.
 We have not systematically studied the possible occurrence of metastable states. But overall our results seem in line with the standard expectation of large N QCD, that mesons (and baryons) are the only bound states, unless some other parameter (in our case the mass hierarchy between quarks and anti quarks) enters the game.

In our analysis we find two types of 4-body bound states with distinct color-coordinate wavefunctions which we refer to as type-I and type-II tetraquarks. In Type-I states the heavy quarks are predominantly in a color anti-symmetric configuration and 
 localized within a region that is much smaller than that where the light anti-quarks are localized.
Instead, in type-II states, the average relative distances among the 
4 constituents are comparable, and moreover
 color and position are strongly entangled. 
Due to the $1/N$ suppression of the BO potentials, the states are parametrically close to the two meson threshold. Moreover for the type-II states we remarkably find a sort of accidental additional closeness to threshold, which originates from the peculiar exponential form of the BO potential.

The type-I tetraquarks are the large $N$ incarnation of states  whose existence was established long ago in QCD for heavy enough quarks \cite{ Carlson:1987hh, Manohar:1992nd}. In color $SU(3)$ these states can be thought of as baryons made of two (anti-)quarks and a tightly bound heavy diquark.
Indeed, following the recent observation of the doubly heavy baryon  $\Xi_{cc}^{++}$\cite{LHCb:2017iph}, its mass has been used to make predictions for tetraquarks containing two heavy quarks and two light anti-quarks, which may be applicable to the observed $T_{cc}^+$. 
This approach was first undertaken in ref.~\cite{Karliner:2017qjm} on the basis of a simple quark model, for which a systematic study of the uncertainties seems unfortunately not possible. In ref.~\cite{Eichten:2017ffp}  a more systematic approach based on heavy quark effective 
theory and quark-diquark symmetry was then undertaken. That    was further significantly refined  in ref.~\cite{Braaten:2020nwp}, which includes also a comprehensive evaluation of  the errors.
 Although these works all agree on the existence below the two meson threshold of tetraquarks containing two $b$ quarks (which are the analogue of our type-I states), they don't agree on tetraquarks containing two charm quarks. 
 More precisely ref.~\cite{Karliner:2017qjm}
 predicts the mass of $T_{cc}$ to be within a few MeV from threshold, while refs.~\cite{Eichten:2017ffp,Braaten:2020nwp} predict it $\mathcal{O}(100)$ MeV above, with a comparably small error.
 But the mass of $T_{cc}$ has in the meantime been measured, and 
it is perhaps as baffling as anything about tetraquarks that the  measured value sits right on threshold, in agreement with the seemingly more qualitative prediction of the quark model, and in disagreement with the prediction of the  more systematic heavy quark EFT approach. On the other hand there is still space for further refinement of the HQEFT analysis, which in its present form neglects effects associated with the finite size of the heavy diquark system.
The leading such correction was already estimated in \cite{An:2018cln} and can significantly affect the mass of $T_{cc}$. Yet another possibility, suggested by our work, is that the observed $T_{cc}$ is more akin to our loosely bound type II tetraquark than to the deeply bound type I.

The existence of stable tetraquarks  at large $N$  with all quark masses large and possibly hierarchical  has been previously studied in \cite{Czarnecki:2017vco}. There, only the states where the two heavier quarks are  bound in a diquark are considered. A  hierarchy of masses is  also found necessary for  the existence of tetraquarks.
However the  condition  they find for the ratio between the quarks and the antiquarks masses is $ M/m\gg N^{3/2}$, which is different and stronger than our $ M/m\gg N$.
We have not been able to sort out the source of disagreement.
On the one hand, in their analytic estimate they require  specific terms to be small for self-consistency,
while we find these terms can be included  in a systematic $1/N$ expansion.  On the other, their   necessary condition for the existence of tetraquarks is  ultimately obtained numerically, making it difficult to find the source of disagreement. 

In recent years the application of the BO approximation to the study of tetraquarks in real world
QCD has started being explored. The grand goal, as outlined for instance in \cite{Braaten:2014qka} would be to use lattice QCD to compute the BO potential among the heavy constituents. Significant progress has then been made in particular for $QQ\bar q\bar q$,
and apparently less so in the case of  $Q\bar Q q\bar q$ (see however \cite{Prelovsek:2019ywc,Sadl:2021pfy}).  In particular \cite{Bicudo:2012qt} computed the potential for two static quarks on the lattice and applied it to the $T_{bb}$ (see ref. \cite{Bicudo:2022cqi} for a recent review of the lattice results) finding bound states below threshold. Other studies, perhaps while waiting for more reliable lattice simulations, have relied on phenomenological modelling of the potential (see e.g.~\cite{Alasiri:2024nue,Maiani:2019cwl}). While these approaches are worthy of consideration, our study of the BO approximation in a fully controllable situation indicates the approximations made by these approaches are probably still too crude. For instance ref \cite{Maiani:2019cwl} works under the assumption of factorized color-coordinate wave functions, while our study shows that the resulting energy eigenstates are often entangled. That is due, as we shall see, to the existence of terms in the Hamiltonian which mix different color singlet configurations and which we can precisely account for.

This paper is organized as follow. In section \ref{sec:Hamiltonian} we write the leading Hamiltonian for the four-quark system and discuss its regime of applicability as well as the main subleading corrections. In section \ref{sec:BO} we study the tetraquark states containing two heavy quarks and two lighter antiquarks using the Born-Oppenheimer approximation, showing the existence of two distinct types of tetraquarks. We also study the excited states and the consequences of the spin-statistics theorem for these tetraquarks. In section \ref{sec:beyondBO} we extend our study beyond regime of applicability of the BO approximation and argue for non-existence of tetraquark ground states in this regime. 
In section \ref{sec:realworld} we discuss to what extent our results may be extrapolated to realistic values of parameters in QCD with $N=3$ and physical quark masses and what they may imply for the tetraquark states.

\section{Hamiltonian} \label{sec:Hamiltonian}
In this section we begin our investigation of the existence of tetraquark states in QCD with a large number of colors $N$ and heavy quark masses by writing the leading Hamiltonian governing the dynamics of the system. We then present a discussion of the subleading corrections which further clarifies the regime of validity of the leading description.
\subsection{The Single  Gluon Exchange Hamiltonian}
A systematic study of the four quark system can be performed in the limit where the quarks are heavy and thus their dynamics is controlled by a non-relativistic Hamiltonian.
At large $N$, the expansion is conveniently organised in terms of $1/N$ and a 't Hooft coupling 
\begin{equation}
    \alpha= \frac{1}{2}\alpha_s N,
\end{equation} 
where $\alpha_s=g_s^2/4\pi$, with $g_s$ being the gauge coupling and the $1/2$ factor is included for later convenience. The strong coupling scale $\Lambda_{\rm QCD}$ is 
the scale at which 
$\alpha$ becomes order unity. 
We work in the regime where all quark masses are heavy,
\beq
m_{i} \gg \Lambda_{\rm QCD}.
\eeq
This implies that $\alpha \ll 1$ evaluated at the relevant scales controlled by $m_i$.
The same parameter $\alpha$ controls gluon emission as well as the relativistic corrections. This can be seen most easily by introducing separate units for space and time and thus reintroducing the speed of light $c$. That way the coupling $\alpha$ is conveniently defined as carrying units of velocity. Higher corrections are then controlled by the dimensionless ratio $\alpha/c$, as systematized within the framework of NRQCD, see e.g. \cite{Manohar_Wise_2000}. The truncation to the non-relativistic Hamiltonian, which we shall employ, is then  self-consistently justified by taking the formal limit $c\to \infty$.

The Hamiltonian of the system of two quarks and two anti-quarks, labeled respectively with indices $1-2$ and $\bar{3}-\bar{4}$, is then given by 
\begin{equation}\label{Energyqqbarx2}
    H = \sum_{i}\frac{p_i^2}{2m_i}+\sum_{i<j}\alpha_s \frac{T^a_{(i)}T^{a}_{(j)}}{r_{ij}}+\text{small corrections}, \hspace{0.6cm}\text{for}\,\, r_{ij}\ll \Lambda_\text{QCD}^{-1},
\end{equation}
where $r_{ij}=|\vec{r}_i-\vec{r}_j|$ are the relative distances between particles $i$ and $j$.
We consider the quarks to be in the fundamental representation of the $SU(N)$ gauge group. The $T^{a}$ matrices are the $N^2-1$ generators of the $SU(N)$ color group in the (anti-) fundamental  
 representation for (anti-) quarks.
The pairwise Coulomb interactions are at most of order $\alpha/ r_{ij} $ (see appendix \ref{app:Wavefunctions} for more details).
Let us note that, because at short distances the running of the 't Hooft coupling is very slow, it is self-consistent to neglect its scale dependence and choose its scale a posteriori as the typical size of the bound state.
Some care has to be taken if the state is characterized by parametrically separated scales.  

The state of two heavy quarks and two heavy anti-quarks can be defined by assigning their position and their color state, as well as 
their flavor and spin. The possible color states come from the tensor product of two fundamental and two anti-fundamental representations of $SU(N)$. This gives rise to two singlets, two adjoints, and four other colored representations given by the tensor product of two adjoints. 
We restrict our analysis to the color singlet subspace, as we expect the ground state to lie in this sector. In the next section we will show
that  is indeed the case, at least for a specific hierarchy of quark masses.
We then write a generic state of the system in the form 
\begin{equation}\label{qqqbarqbarstate}
    \ket{\Psi}=\sum_{\rho} \int \prod_{k=1}^4 d^3 r_k \, \Psi^{i\,j\,}_{m \,n\,} (r,\rho)\ket{1_i(r_1,\rho_1)\,2_j(r_2,\rho_2)\,\bar{3}^m(r_{\bar{3}},\rho_{\bar{3}})\,\bar{4}^n(r_{\bar{4}},\rho_{\bar{4}})},
\end{equation}
with $\Psi^{i\,j\,}_{m \,n\,} $ invariant under the action of $SU(N)$ on the color indices ($i,j,m,n$). We have also collectively denoted the flavor and spin quantum numbers by $\rho$. The wave function must be localized inside the region $r_{ij}\ll \Lambda_\text{QCD}^{-1}$ for its dynamics to be controlled by the Hamiltonian in (\ref{Energyqqbarx2}). As there exist two independent color singlet contractions of the four color indices, the wave function spans a two-dimensional  subspace. 
Different choices for the basis of this subspace can be made, with their convenience depending on the question being asked and the regime being considered.
(for a more detailed exposition see Appendix \ref{app:Wavefunctions}). One possibility is to pick  a basis where one element corresponds to a pair of $q\bar{q}$ singlets, while the orthogonal element  corresponds to a state  where the same $q\bar{q}$ pairs lie in the adjoint representation. Obviously, there exist two such options, corresponding to the two possible pairings, either $1\bar{3}$ and $2\bar{4}$ or $1\bar{4}$ and $2\bar{3}$.  A more ``symmetric'' basis is obtained by first considering the two states  where the $qq$ lie in either a color symmetric ($\Psi_S$) or anti-symmetric ($\Psi_A$) configuration, with the $\bar{q}\bar{q}$ pair  in the conjugate representation so as to make up a singlet, and by then forming the combinations
\begin{equation}\label{plusminusbasis}
    \Psi_+=\frac{1}{\sqrt{2}}(\Psi_S+\Psi_A), \hspace{1cm} \Psi_-=\frac{1}{\sqrt{2}}(\Psi_S-\Psi_A).\\
\end{equation}
 In color space the potential is then given by a two dimensional matrix,
\begin{equation}
    V = \begin{pmatrix}
 V_{++} \;  & V_{-+} \\ V_{+-} \; &  V_{--}
  \end{pmatrix}
\end{equation}
with elements (see Appendix \ref{app:potential})
\begin{equation}\label{potentialplusminus}
\begin{split}
    &V_{++}=-\frac{\alpha}{r_{1\bar{3}}}-\frac{\alpha}{r_{2\bar{4}}}+\mathcal{O}\bigg(\frac{1}{N^2}\bigg),\\
    & V_{+-}=V_{-+}=\frac{\alpha}{2N}\left(\frac{2}{r_{12}}+\frac{2}{r_{\bar{3}\bar{4}}}-\frac{1}{r_{1\bar{3}}}-\frac{1}{r_{1\bar{4}}}-\frac{1}{r_{2\bar{3}}}-\frac{1}{r_{2\bar{4}}} \right)+\mathcal{O}\bigg(\frac{1}{N^2}\bigg),\\
    &V_{--}=-\frac{\alpha}{r_{1\bar{4}}}-\frac{\alpha}{r_{2\bar{3}}}+\mathcal{O}\bigg(\frac{1}{N^2}\bigg).\\
\end{split}
\end{equation}
The $N\rightarrow \infty$ limit, with the quark masses kept fixed, is manifest. The mixing between the singlets vanishes and the diagonal elements consists of just two $q\bar{q}$ Coulombic potentials. In this limit, 
\begin{equation}
\begin{split}
       &\Psi_+\rightarrow(1\bar{3})_{\text{singlet}}(2\bar{4})_{\text{singlet}},\\
       &\Psi_-\rightarrow(1\bar{4})_{\text{singlet}}(2\bar{3})_{\text{singlet}},
\end{split}
\end{equation}
and the spectrum corresponds to that of two free mesons for both possible pairings. When $N$ is large but finite, the physics of this system is richer. In particular, we will show that the $1/N$ corrections can form tetraquark states in specific regimes of the particles masses. 
Since in the Hamiltonian in eq.~\eqref{Energyqqbarx2}, we keep the $1/N$ corrections and neglect the $\alpha^2$ ones, it naively seems to be necessary to impose  $\alpha\ll 1/N$. However, as we will show below, this is not the case. 
The leading interactions in $1/N$, to any order in $\alpha$,
only modify the Coulombic interactions among the pairs of quark anti-quark binding into mesons when $N\rightarrow\infty$. Thus they do not give rise to interactions among the two pairs. 
Indeed, they correspond to diagrams with the two-meson topology that is two fermion loops that must be disconnected in order to survive in the $N\rightarrow \infty$ limit. A more detailed large $N$ counting is provided in the next section. For our purposes, the knowledge of the $q\bar{q}$ interactions at leading order in $\alpha$ will be sufficient to compute the distance from threshold of the tetraquark states to first order in $\alpha$ and in $1/N$.

We will then be interested in the study of the bound states of this system as a function of the particle masses in the region where we can gain some analytic understanding. To this purpose, the Hamiltonian previously defined is too complicated as it generically entails the solution of a four-body problem. The dynamics can be simplified if we consider the regime with a mass hierarchy between the four particles. More specifically, we will study the situation where two of them are heavier, with masses of order $M$ while the other two have a mass of order $m$ with $\Lambda_\text{QCD}\ll m\ll M$.  Up to charge conjugation, there are two classes of systems, that where two quarks are heavy, denoted as $QQ\bar{q}\bar{q}$, with masses $M_1, M_2 =\mathcal{O}(M)$ and $m_{\bar{3}},m_{\bar{4}} =\mathcal{O}(m)$, and that where the heavy particles are a quark and an anti-quark $Q\Bar{Q}q\bar{q}$ with $M_1, M_{\bar{3}} =\mathcal{O}(M)$ and $m_2,m_{\bar{4}} =\mathcal{O}(m)$. 
The hierarchy $m/M\ll 1$ does not guarantee by itself the separation between two dynamical scales as these are determined by the structure of the interactions and the reduced masses of the system. Indeed only for the $QQ\bar{q}\bar{q}$ case do we find self-consistent bound states in an expansion in powers of $m/M$. At leading order, one can first solve for the dynamics of the light particles with the heavy ones providing a static background and then use the solution to generate an effective potential for the heavy quarks. 
This is the Born-Oppenheimer approximation which will be shown to be valid as long as $m/M < 1/N$.

As the mass hierarchy becomes larger $m/M  \lesssim 1/N^2$, the heavy quarks eventually dominate the binding energy of the lowest-lying states and their dynamic becomes faster than that of the anti-quarks. For this reason, it is convenient to use a basis of states with a definite color configuration of the quarks. These can either be in a symmetric or an anti-symmetric configuration. The matrix elements of the potential in this basis are

\begin{equation} \label{eq:Potential_4quarks_SA}
\begin{split}
     &V_{SS}= - \frac{\alpha}{2}\left(\frac{1}{r_{1\bar{3}}}+\frac{1}{r_{1\bar{4}}}+\frac{1}{r_{2\bar{3}}}+\frac{1}{r_{2\bar{4}}} \right)\\
     &\hspace{1.1cm}+\frac{\alpha}{2N}\left(\frac{2}{r_{12}}+\frac{2}{r_{\bar{3}\bar{4}}}-\frac{1}{r_{1\bar{3}}}-\frac{1}{r_{1\bar{4}}}-\frac{1}{r_{2\bar{3}}}-\frac{1}{r_{2\bar{4}}} \right)+\mathcal{O}\bigg(\frac{1}{N^2}\bigg),   \\
     &V_{SA}=V_{AS}=- \frac{\alpha}{2}\left(\frac{1}{r_{1\bar{3}}}+\frac{1}{r_{2\bar{4}}}-\frac{1}{r_{1\bar{4}}} -\frac{1}{r_{2\bar{3}}}\right)+\mathcal{O}\bigg(\frac{1}{N^2}\bigg), \\
     &V_{AA}=- \frac{\alpha}{2}\left(\frac{1}{r_{1\bar{3}}}+\frac{1}{r_{1\bar{4}}}+\frac{1}{r_{2\bar{3}}}+\frac{1}{r_{2\bar{4}}} \right)\\
     &\hspace{1.1cm}-\frac{\alpha}{2N}\left(\frac{2}{r_{12}}+\frac{2}{r_{\bar{3}\bar{4}}}-\frac{1}{r_{1\bar{3}}}-\frac{1}{r_{1\bar{4}}}-\frac{1}{r_{2\bar{3}}}-\frac{1}{r_{2\bar{4}}} \right)+\mathcal{O}\bigg(\frac{1}{N^2}\bigg). 
\end{split}
\end{equation}
In this regime, one can thus solve first for the states of the heavy pair, which bind at short distances in an anti-symmetric state, and then consider the system of the compact di-quark interacting with the two anti-quarks.

\subsection{Corrections to the Single Gluon Exchange Hamiltonian}

In this subsection we discuss why we can include the interactions of order $\alpha/N$ in eq.~\eqref{Energyqqbarx2} while dropping terms of $\mathcal{O}(\alpha^2)$ 
without assuming a hierarchy between the 't Hooft coupling $\alpha$ and $1/N$. We also justify why we can limit our analysis to the singlet subspace. 
The reader who is satisfied with these statements  can directly skip to section \ref{sec:BO}. 

\begin{figure}
    \centering
    \includegraphics[scale=0.35]{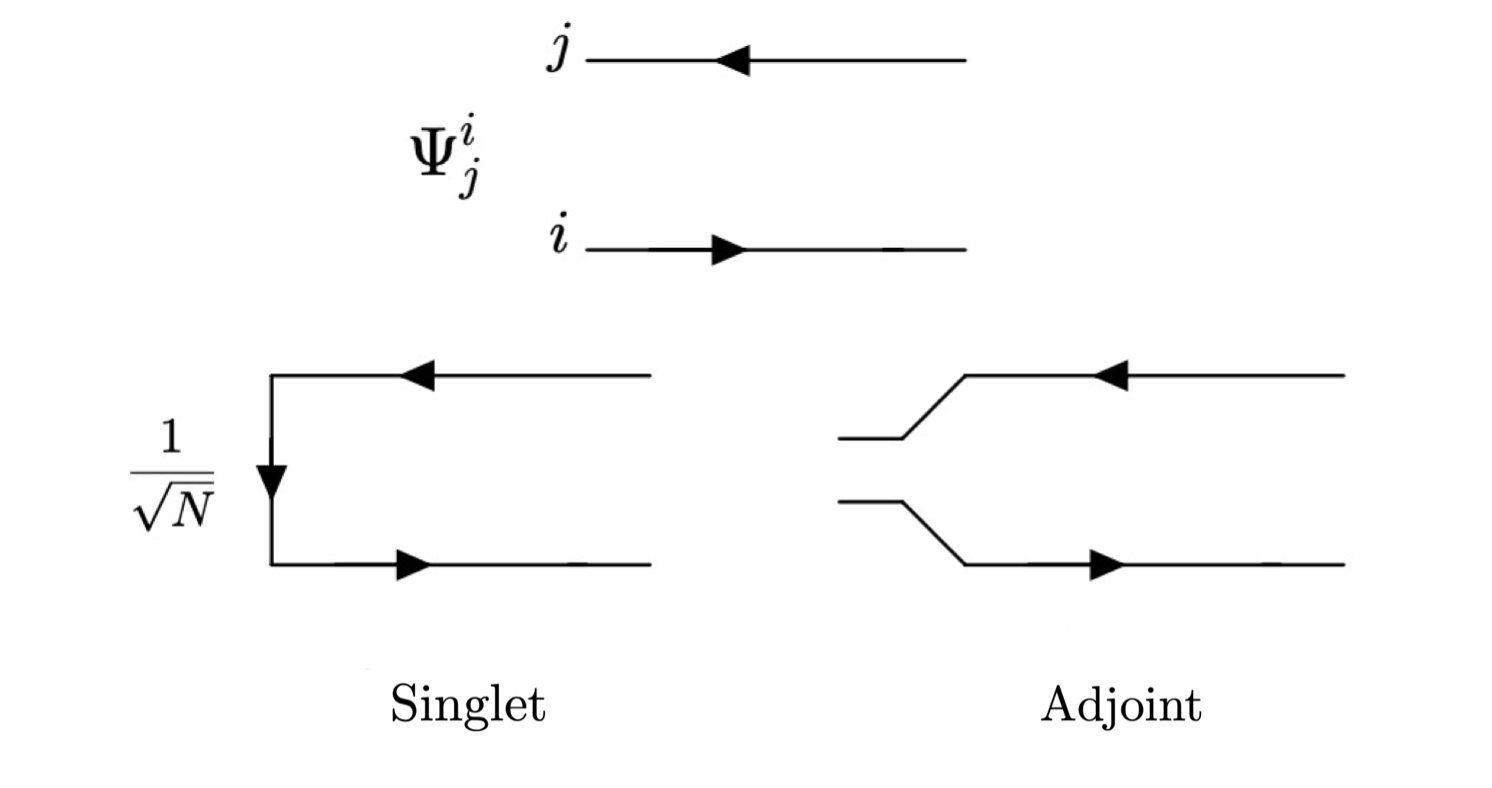}
    \caption{Diagrammatic representation of the color wave functions of a $q\bar{q} $ system. The open color line represents an adjoint index labeling the possible $N^2-1$ states in the representation.
    }
    \label{fig:WF}
\end{figure}

To power count the different contributions to the Hamiltonian of the system, we study the position space propagator of the two quarks and two anti-quarks. 
This is expanded in diagrams with four incoming and four outgoing fermion lines each of which carries either a fundamental or an anti-fundamental color index. These indices will be contracted with the ones of the wave functions of the possible color states of the quarks and anti-quarks. To proceed in the usual counting of powers of $N$, we thus need to give a diagrammatic representation for the external states. This is easily done once the color wave functions are known. Indeed, they are constructed in terms of Kronecker deltas $\delta^{i}_j$, and the generators of the fundamental representation $(T^{a})^i_j$ for which the double line notation is the canonical one used in large $N$. As an example, consider the color state of a $q\bar{q}$ pair, this can be either a singlet or one of the $N^2-1$ states of the adjoint representation. The wave functions are given by $\frac{1}{\sqrt{N}}\delta^{i}_j$ and $\sqrt{2}(T^{a})^i_j$ respectively. The diagrammatic contraction with a quark and an anti-quark line is represented in figure \ref{fig:WF}. 
\\

As for the system of two quarks and two anti-quarks, the tensor product of two fundamental and two anti-fundamental representations in $SU(N)$ gives rise to two singlets, two adjoints, and four other colored representations coming from the tensor product of two adjoints. We start studying the order of the corrections
within the singlet subspace, 
before studying the mixing with higher dimensional representations in the next subsection. For the ease of the reader, let us recall some of the results in the large $N$ counting that will use in the following.
\begin{itemize}
    \item The leading contribution in $1/N$ is a sum of diagrams whose boundary is defined by fermion lines and planar gluons decorate its interior. 
    \item Non planar gluon corrections come in powers of $1/N^2$.
    \item Internal quark loops are suppressed by $1/N$ with respect to a gluon loop with the same topology. For this reason, they can be neglected. 
\end{itemize} 

\subsubsection{Singlet subspace}
\begin{figure}
    \centering
    \includegraphics[width=0.90\linewidth]{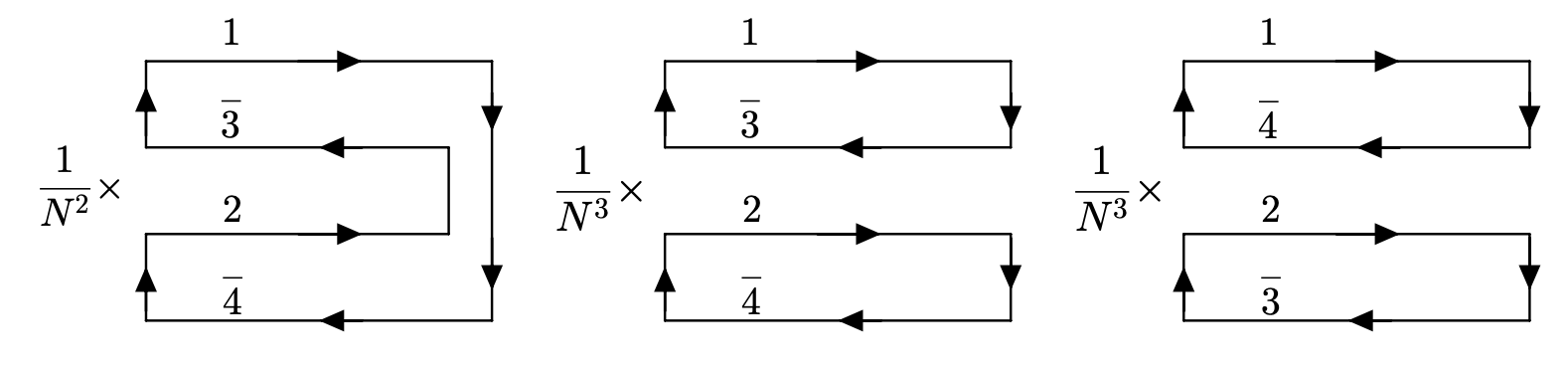}
    \caption{Structure of the leading diagrams contributing to the off-diagonal entries of the Hamiltonian in the basis (\ref{plusminusbasis}). The wave function factor ($1/N^2$ for the first diagram and $1/N^3$ for the other two diagrams) combines with the factors coming from the fermion loops ($N$ for the first diagram, $N^2$ for the other two) to give the term in equation (\ref{potentialplusminus}). }
    \label{fig:offdiagonaldiagram}
\end{figure}

Let us start studying the Hamiltonian in the singlet subspace. The wave-functions for the states defined in (\ref{plusminusbasis}) to sub-leading order in $1/N$ are (see appendix \ref{app:Wavefunctions})
\begin{equation}\label{wfpmlargeN}
\begin{split}
    &P(+)^{ij}_{mn}\equiv P(S)^{ij}_{mn}+P(A)^{ij}_{mn}=\frac{\sqrt{2}}{N} \delta^i_m\delta^j_n+\frac{1}{\sqrt{2}N^2} \delta^i_n\delta^j_m + \mathcal{O}\bigg(\frac{1}{N^3}\bigg),\\
    &P(-)^{ij}_{mn}\equiv P(S)^{ij}_{mn}-P(A)^{ij}_{mn}=\frac{\sqrt{2}}{N} \delta^i_n\delta^j_m +\frac{1}{\sqrt{2}N^2} \delta^i_m\delta^j_n + \mathcal{O}\bigg(\frac{1}{N^3}\bigg). 
\end{split}
\end{equation}
\begin{figure}
    \centering
    \includegraphics[scale=0.14]{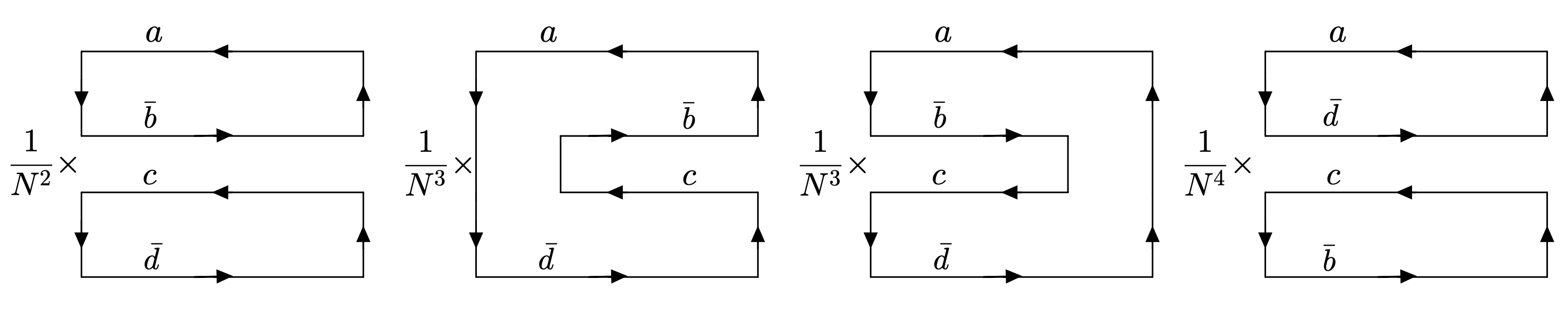}
    \caption{Structure of the fermion lines of the diagrams contributing to the diagonal entries of the Hamiltonian in the basis (\ref{plusminusbasis}). The explicit factors of $1/N$ come from the wave functions (\ref{wfpmlargeN}). The letters denote which fermion is related to that line of the propagator. For the $++$ component we have $a=1,\bar{b}=3,c=2,\bar{d}=4$ while $a=1,\bar{b}=4,c=2,\bar{d}=3$ for the $--$ one. When decorated with gluons, each diagram gives at most a contribution of the order of the wave-function prefactor multiplied by $N$ to the power of the number of fermion loops.}
    \label{fig:diagonaldiag}
\end{figure}
As shown in figure \ref{fig:diagonaldiag}, we see that the diagrams contributing to the diagonal elements of the Hamiltonian, at leading order in $1/N$ , are made of two fermion loops. Other structures of fermion lines are suppressed at least by $1/N^2$. This effect comes either from the wave function factor (as in the rightmost diagram of figure \ref{fig:diagonaldiag}) or from the combined contribution of the topology of the diagram and the wave function (as in the middle diagrams of the figure).  The two fermion loops must then be decorated with gluons in all possible ways. There are two types of decorations. The ones connecting the loops and the ones that don't. One diagram of each type is shown in figure \ref{fig:gluoncorrections}. The former, besides additional powers of the 't Hooft coupling, are $1/N^2$ suppressed. The latter, on the contrary, give rise to $\alpha$ corrections to the SGE Hamiltonian at leading order in $1/N$. However, they all share the structure of a two meson state and they will not generate interactions between the mesons that can compete with the off-diagonal ones at order $\alpha/N$. We then conclude that the corrections to the diagonal elements of the single gluon exchange Hamiltonian that give rise to interactions among the mesons come at order $1/N^2$. As regards the off-diagonal element, the one in (\ref{potentialplusminus}) is the leading one. Indeed, the dominant diagrams in $1/N$ are shown in figure \ref{fig:offdiagonaldiagram}.
\begin{figure}
    \centering
    \includegraphics[scale=0.3]{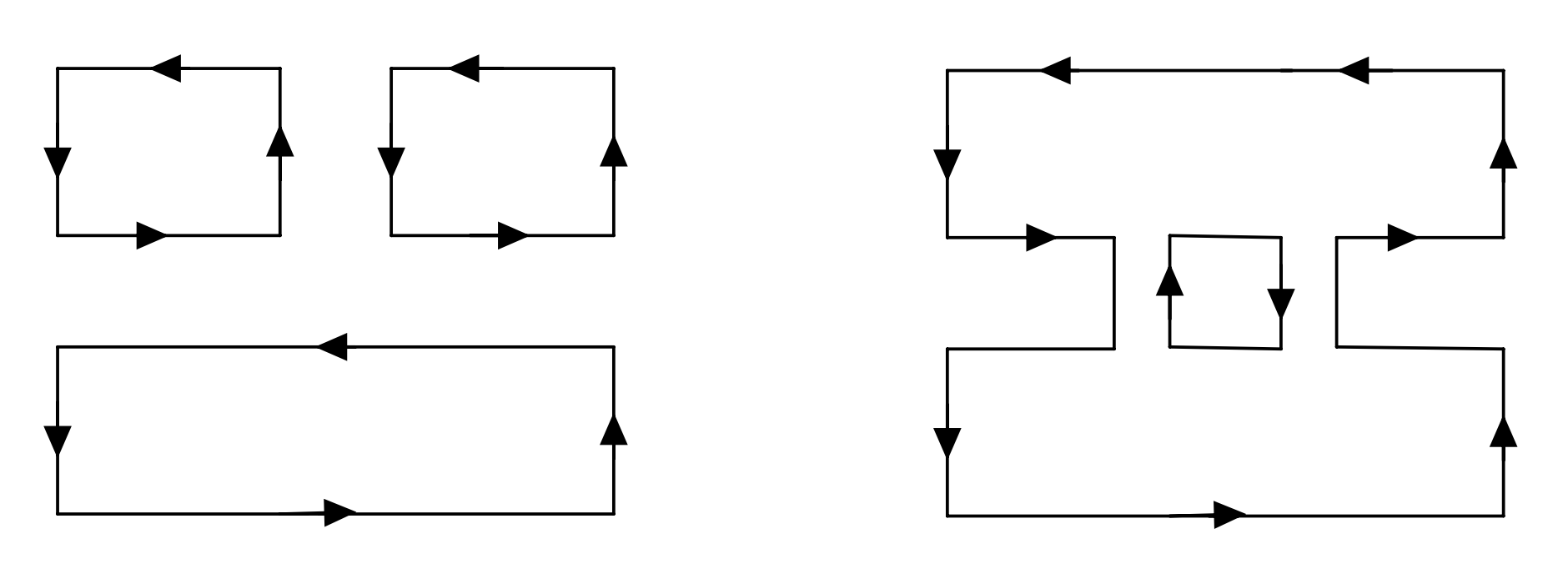}
    \caption{Examples of planar (left) and non planar (right) gluon corrections to the first diagram in Figure \ref{fig:diagonaldiag}. The diagram on the left contribute at order $\alpha$ while the one on the right gives a contribution of order $\alpha^2/N^2$. }
    \label{fig:gluoncorrections}
\end{figure}

\subsubsection{Mixing with higher dimensional representations}
As we stated before, we expect the ground state of the system to be dominantly in the singlet subspace. In some special cases, this is easy to see. For example in the mass hierarchy $M_{1,2}\gg N^2 m_{\bar{3}}\gg N^2 m_{\bar{4}}$, the problem reduces to a series of two-body problems that we can easily analyze. The leading order problem consists of two heavy quarks, and the binding energy is maximal if the two heavy quarks forming a compact color anti-symmetric diquark. Then including the lighter anti-quarks one-by-one we can see that the configuration maximizing the binding energy is a total color-singlet. A similar conclusion is found for
$N^2 m_{\bar 4} \gg M_{1,2} \gg N m_{\bar 3} \gg N m_{\bar 4} $. This time the binding energy of the full system is dominated by forming the color singlet meson involving $\bar{q}_{\bar{3}}$, while the leading corrections come from forming the meson involving $\bar{q}_{\bar{4}}$. The full system is therefore in the color singlet subspace, up to small corrections. From now on we assume that the ground state is dominantly a color singlet, and investigate the mixing with the other color representations.

Besides the singlets, the two quarks and two anti-quarks can lie in higher dimensional representations where the color is neutralized by additional gluons. The tensor product $N\otimes N\otimes\bar{N}\otimes \bar{N}$ gives rise to two adjoint representations and four other irreducible representations whose color must be screened by at least two gluons. If the color is neutralized at the length scale $\Lambda_\text{QCD}^{-1}$, we expect any mixing to be suppressed by powers of $\Lambda_\text{QCD}/\alpha m$ with $m$ denoting collectively the mass of the quarks. However, if the gluons can localize at a much shorter scale, binding the quarks with the anti-quarks, we expect the mixing to be suppressed only by powers of the weak coupling $\alpha(\text{binding scale})^{1/2}$. At least one power is needed for the adjoint states while two are needed for the others. Nevertheless corrections that survive as $N\rightarrow \infty$, can only give rise to interactions that modify the Coulombic potential between a quark/anti-quark pair. This stems from the fact that the topology of the diagrams associated with interactions among the ``mesons" necessarily corresponds to sub-leading order in $1/N$, just as above for the corrections within the singlet sector.
Said differently, they only modify the meson states of the $N\rightarrow \infty$ Hamiltonian mixing the $q\bar{q}$ singlet with $q\bar{q}+\text{gluons}$ at some sub-leading order in $\alpha$. Therefore, for the purpose of determining the leading 
interaction among the ``mesons", 
it is sufficient to consider the singlet subspace.

\section{Tetraquarks within the Born-Oppenheimer approximation: two heavy quarks and two lighter antiquarks}\label{sec:BO}

In this section, we begin our analysis of the Hamiltonian of eq.~\eqref{Energyqqbarx2} focussing on a specific mass hierarchy, where the quarks are  much heavier than the antiquarks. Denoting  the masses of the quarks by $M_1$ and $M_2$ and those of the antiquarks by $m_{\bar{3}}$ and $m_{\bar{4}}$, our starting assumption is then \footnote{The  case where the antiquarks are much heavier than the quarks is simply related to this one by charge conjugation.}
\begin{equation}
    M_1 \geq M_2 \gg m_{\bar{3}} \geq m_{\bar{4}} \gg \Lambda_\text{QCD}\,.\label{hierarchy}
\end{equation}
 As it will become clear below, it is convenient to introduce the following coordinates
\begin{align}
    \vec{R}_{CM} &= \frac{M_1\vec{r}_1+M_2\vec{r}_2+m_{\bar{3}} \, \vec{r}_{\bar 3}+m_{\bar{4}}\, \vec{r}_{\bar 4}}{M_1+M_2+m_{\bar{3}}+m_{\bar{4}}} ,\\
    \vec{R} &= \vec{r}_2-\vec{r}_1 ,\\
    \vec{r}^{\, '}_{\bar{3}}&= \vec{r}_{\bar{3}} - \frac{\vec{r}_1+\vec{r}_2}{2} ,\label{r3prime}\\
    \vec{r}^{\, '}_{\bar{4}}&=\vec{r}_{\bar{4}} - \frac{\vec{r}_1+\vec{r}_2}{2},\label{r4prime}
\end{align}
with their corresponding momenta denoted by $\vec{P}_{CM}$, $\vec{P}$, $\vec{p}^{\; '}_{\bar{3}}$, and $\vec{p}^{\; '}_{\bar{4}}$. 
Galilean  invariance ensures  the decoupling of the dynamics of the center of mass (CM) canonical pair $(\vec R_{CM}\,,\vec P_{CM})$.  For  the bound state problem we then need to  consider only $\vec R$, $\vec r^{\, '}_{\bar 3}$, $\vec r^{\, '}_{\bar 4}$ and their conjugated momenta. Notice that  $\vec{r}^{\, '}_{\bar{3}/\bar{4}}$ 
are simply the  distances of the light anti-quarks  from the midpoint of quark 1 and 2, which can be interpreted as a sort of center of color charge. This choice has been made for later convenience.
In these coordinates, the Hamiltonian reads
\begin{equation} \label{eq:Hamiltonian_after_coordinate_trafo}
   H = \frac{P_{CM}^2}{2 \left(M_1+M_2+m_{\bar{3}}+m_{\bar{4}}\right)}+  
    \frac{P^2}{2 M_{12}}+
    \frac{p'^2_{\bar{3}}}{2 m_{\bar{3}}}+
    \frac{p'^2_{\bar{4}}}{2 m_{\bar{4}}} + V + \text{corrections}. 
\end{equation}
with $M_{12}\equiv M_1M_2/(M_1+M_2)$  the reduced mass of the heavy quark system. The corrections not written explicitly above consist of terms of the form
    $\frac{P p'_i}{M} $ and $ \frac{p'_j p'_i}{M}$, where $M$ is a heavy quark mass. We will see below that within the Born-Oppenheimer  approximation these terms can be consistently dropped.

To apply the Born-Oppenheimer approximation (see e.g. \cite{weinberg_2015} and appendix \ref{app:BOH2}), we first focus on a reduced Hamiltonian for the light antiquarks
\begin{equation}
    H_R =  \frac{p'^2_{\bar{3}}}{2 m_{\bar{3}}}+
    \frac{p'^2_{\bar{4}}}{2 m_{\bar{4}}} +V\,,
    \label{HR}
\end{equation}
where we neglect all the (kinetic) terms suppressed by the heavy quark masses, and where we treat $\vec R$, which appears in  $V$, as a classical parameter.

The energy eigenvalues and eigenstates of the reduced Hamiltonian, satisfying 
\begin{equation}
    H_R |\psi_{\cal A}\rangle = E_{\cal A} |\psi_{\cal A}\rangle\,,
\end{equation}
with ${\cal A}$ a collective quantum number,
can then be found working in a $1/N$ expansion.
The eigenvalues $E_{\cal A}$, with their dependence on $\vec R$, then provide the BO potential for the $Q_1{\text -}Q_2$ system. More precisely, each 
light quark state $|\psi_{\cal A}\rangle$, leads to an approximate effective Hamiltonian
\begin{equation}
\label{BOhamiltonian}
    H_{BO}=\frac{ P^2}{2 M_{12}}+E_{\cal A}( R)
\end{equation}
whose Schr\"odinger equation provides the  approximate wavefunctions and energy levels of the bound 4-quark system. Notice that $H_R$ is invariant under rotations when treating the parameter $\vec R$
as a vector. Therefore its eigenvalues can only be functions of the norm $|\vec R|\equiv R$, corresponding to a spherically symmetric BO potential.

A crucial final step is to
check for the self-consistency of the BO approximation. A simple example of the BO approximation for a molecule with large charge nuclei, where a systematic analysis can be performed analytically, is presented in appendix \ref{app:BOH2explicit}. 
The approximation is valid if the motion of the heavy quarks has negligible influence on the wavefunction of the light antiquarks. As also reviewed in appendix \ref{app:BOH2explicit}, that happens to be  the case when the  heavier quarks are much more localized than the lighter antiquarks, or, equivalently, in terms of their momenta (see eq.~(\ref{eq:Hamiltonian_after_coordinate_trafo}))
\begin{equation}\label{eq:conditionBO}
    P \,\gg \,p'_{\bar{3},\bar{4}}.
    \end{equation}
   That is also the same condition that allows us to drop the corrections to the kinetic terms in eq.~\eqref{eq:Hamiltonian_after_coordinate_trafo}. 
As we will show, it reduces in our case  to a condition on the masses:
\beq\label{eq:conditionBOmass}
M_2\gg N m_{\bar{3}}.
\eeq 
The scaling with $N$ comes from the fact that the BO potential is only generated at subleading order in $1/N$, while at leading order, the energy eigenstates are a set of approximately color-singlet free ``mesons".

In the regime of eq.~\eqref{eq:conditionBOmass}, we find two distinct tetraquark solutions, while 
in the regime $m_{\bar{3}} \ll M_2 \ll N m_{\bar{3}}$ we show 
that there are no tetraquark bound states within the domain of validity of the BO approximation. In section \ref{sec:beyondBO}, we offer  a general argument indicating that in this other regime the ground state is a two meson state and not a tetraquark.

\subsection{Leading order in $1/N$: The mesons}

In this subsection,  we study  the reduced $H_R$ Hamiltonian of eq.~\eqref{HR}, at leading order in $1/N$. In the $+$/$-$ basis of eq.~\eqref{potentialplusminus} the potential matrix then becomes
\beq \label{eq:Leading_Potential_pm}
 V=\alpha
\begin{pmatrix}
 - \frac{1}{r_{1\bar{3}}} - \frac{1}{r_{2\bar{4}}}   & 0 \\ ~\\
0 &  - \frac{1}{r_{1\bar{4}}} - \frac{1}{r_{2\bar{3}}}
  \end{pmatrix} + \mathcal{O}(\alpha/N).
\eeq

The Hamiltonian with the leading order potential is straightforward to solve and simply gives rise to two pairs of free ``mesons":  $(Q_1 \bar{q}_{\bar{3}})$ and $(Q_2 \bar{q}_{\bar{4}})$  with the $+$ color configuration and $(Q_2 \bar{q}_{\bar{3}})$ and $(Q_1 \bar{q}_{\bar{4}})$ in the $-$ configuration.
Indeed the $+$ ($-$) states, up to $1/N$ corrections, correspond to configurations where  $Q_1$ and  $Q_2$ form singlets with $\bar{q}_{\bar{3}}$ ($\bar{q}_{\bar{4}}$) and $\bar{q}_{\bar{4}}$ ($\bar{q}_{\bar{3}}$) respectively. 
 $H_R$ has then two degenerate ground states, corresponding to the two different meson pairs. Their energy $E_0$ is simply given by
\beq
E_0 = -\mathcal{E}_{\bar{3}} -\mathcal{E}_{\bar{4}}  \qquad \text{with} \qquad  \mathcal{E}_{\bar{3}} = \frac{1}{2}\alpha^2 m_{\bar{3}}  \qquad \text{and}  \qquad \mathcal{E}_{\bar{4}} = \frac{1}{2}\alpha^2 m_{\bar{4}},
\eeq
with $\mathcal{E}_{\bar{3}}$ and $\mathcal{E}_{\bar{4}}$ respectively the binding energies of the meson involving $\bar{q}_{\bar{3}}$ and  $\bar{q}_{\bar{4}}$. The mesons have Bohr radii 
\beq a_{\bar{3}}= (\alpha m_{\bar{3}})^{-1}, \qquad \text{and} \qquad a_{\bar{4}}= (\alpha m_{\bar{4}})^{-1}.
\eeq 

Note that in the BO approximation, the heavy quarks $Q_1$ and $Q_2$ are treated as static, 
and thus in the above equations it is indeed $m_{\bar{3}}$ and $m_{\bar{4}}$ that appears and not the respective reduced masses.  
It is clear that within the BO approximation, at this order in $1/N$, the energy eigenvalues are independent of the position of the heavy quarks and no BO potential is generated. Moreover, at this order there is a degeneracy between the energy eigenvalues in + and - color configurations. As we shall now see, this degeneracy is broken by the  $1/N$ corrections  which also gives rise to a BO-potential that can bound the heavy mesons together.

\subsection{Subleading in $1/N$: The Born-Oppenheimer potential}
\begin{figure}[h!]
    \centering
\includegraphics[width=.7\linewidth]{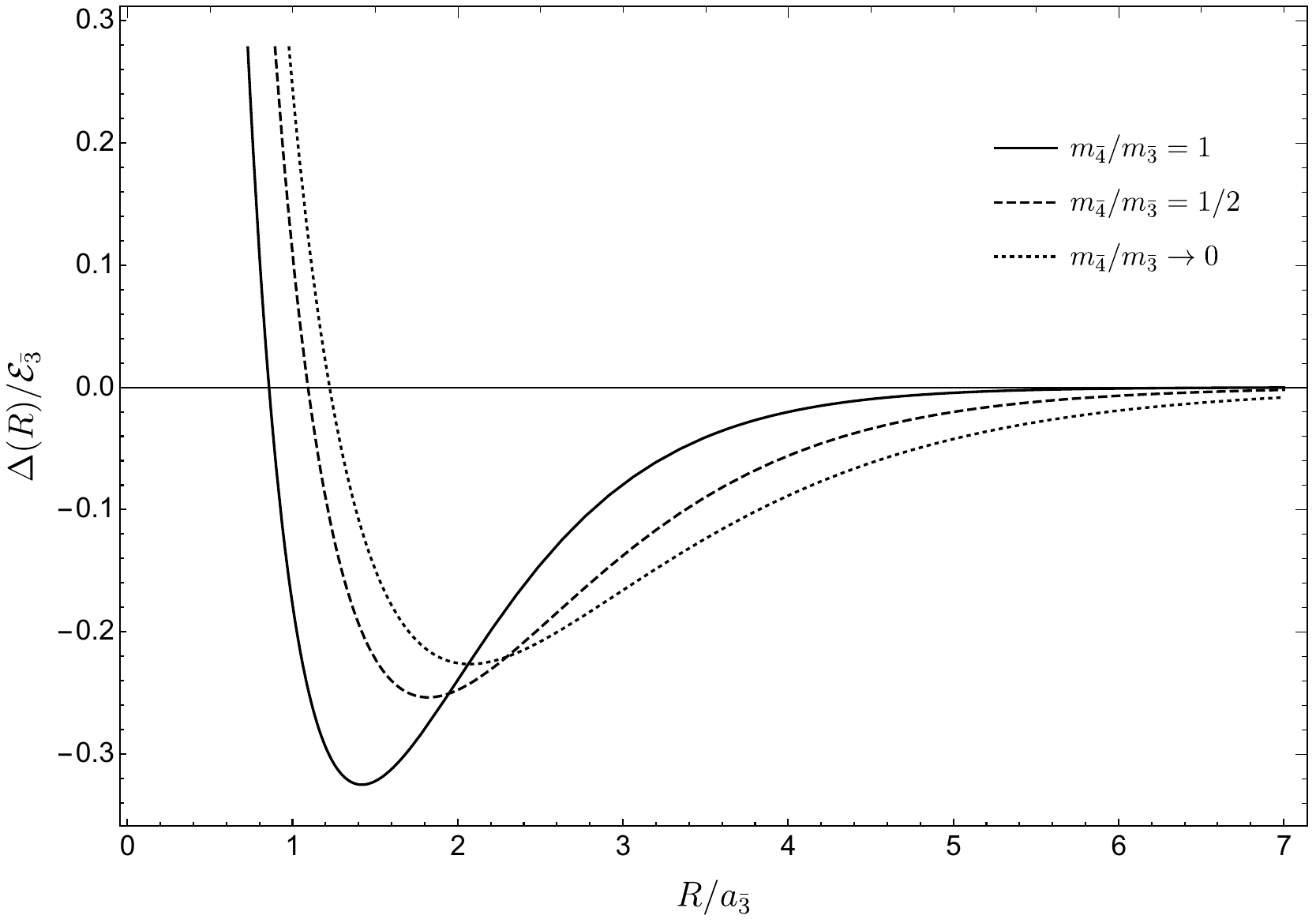}
    \caption{ $\Delta (R)/\mathcal{E}_{\bar{3}}$ for different values of $ 
    m_{\bar{4}}/m_{\bar{3}} $. One can see the minimum at $R\sim O(a_{\bar{3}})$, which leads to type-II tetraquarks, as well as the $1/R$ behaviour at small distances that gives rise to type-I tetraquarks.} 
    \label{fig:BO_4quarks_eta}
\end{figure}

According to the discussion in the previous subsection,
at leading order in $1/N$ the ground state of the lighter quark dynamics is independent of $R\equiv r_{12}$ and has a two-fold degeneracy. We denote the two ground states by $|{\psi_0^\pm}\rangle$, with $\pm$ superscript specifying their color configuration. The subleading $1/N$ effects
split the degeneracy by an $R$-dependent correction, with the resulting ground state 
being  a linear combination of the two initially degenerate states.
In order to study that, we first compute the matrix element of the potential between the two (degenerate) leading order ground states,
\beq \label{eq:defineDelta}
\frac{\Delta(R)}{N}= \langle{\psi^+_0}|V_{+-} |{\psi_0^-}\rangle,
\eeq
where we  factored out a $1/N$ so that $\Delta(R)$ does not scale with $N$.  In terms of this matrix element and of  the the leading order ground state energy, $E_0=-\mathcal{E}_{\bar{3}}-\mathcal{E}_{\bar{4}}$, the energy eigenvalues are $E_0\pm \frac{\Delta(R)}{N}$ and correspond to the states $\ket{\psi^+_0}\pm \ket{\psi^-_0}$.  The BO potentials (with the free meson energies subtracted) for  $\ket{\psi^+_0}\pm \ket{\psi^-_0}$ are therefore simply given by $V^\pm_{\rm eff,BO }(R)=\pm\frac{\Delta(R)}{N}$. In figure \ref{fig:BO_4quarks_eta}, we show $\Delta(R)$ for various choices of $
 \frac{m_{\bar{4}}}{m_{\bar{3}}}$. In the limit $\frac{m_{\bar{4}}}{m_{\bar{3}}} \rightarrow 0$, it takes the following simple analytic form
\beq\label{eq:potentialequalmass}
\frac{\Delta(R)}{ \mathcal{E}_{\bar{3}}}\Big|_{m_{\bar{4}}/m_{\bar{3}} \rightarrow 0}= 2 e^{-R/a_{\bar{3}}} \left( \frac{a_{\bar{3}}}{R} - \frac{2}{3} \frac{R}{a_{\bar{3}}} \right).
\eeq
For $R \ll a_{\bar{3}}$ this is well approximated by a repulsive $\propto \frac{1}{R}$ potential, clearly resulting from the $\frac{1}{r_{12}}$ interaction between $Q_{1}$ and $Q_2$. At $R\gg a_{\bar{3}}$, the overlap of the spatial wavefunction of the states is exponentially suppressed and so is $\Delta(R)$. 
These asymptotic behaviors also hold for generic $\frac{m_{\bar{4}}}{m_{\bar{3}}}$. 
The dependence of the curves on $\frac{m_{\bar{4}}}{m_{\bar{3}}}$ can also be understood as follows. As we increase $\frac{m_{\bar{4}}}{m_{\bar{3}}}$,
 the overlap $|\Delta (R)|$ drops faster at large $R$ because of the more spacially localized  $\bar{q}_{\bar{4}}$ wavefunction. On the other hand,  at small $R \lesssim a_{\bar{3}}$, the same increased localization of $\bar{q}_{\bar{4}}$ boosts the negative contribution to $\Delta(R)$ of the  terms proportional to $\frac{1}{r_{1\bar{4}}}$ and $\frac{1}{r_{2\bar{4}}}$ in $V_{+-}$ (see eq.~(\ref{potentialplusminus})), thus leading to a smaller $\Delta(R)$. The analytic expression for $\Delta(R)$ expanded up to second order for  $m_{\bar{4}}/m_{\bar{3}} \ll 1$ is given in the appendix \ref{app:BOpot}.  For  $m_{\bar{4}}/m_{\bar{3}} \ll 1$, $\Delta(R)$  is dominated by the contribution from $m_{\bar{3}}$, so that the results become independent of $m_{\bar{4}}$. In particular  they are unaffected by  $m_{\bar{4}}$  being  bigger or smaller than $\Lambda_\text{QCD}$.

We can easily see from figure~\ref{fig:BO_4quarks_eta}, that the BO potential can potentially give rise to two distinct tetraquark bound states:  one on the $(-)$ branch  where $V^-_{\rm eff,BO}=-\frac{\Delta(R)}{N}$ provides   at $R \ll a_{\bar{3}}$ an attractive $\propto \frac{1}{R}$ potential that can localize the two heavy quarks, and the other on the $(+)$ branch where $V^+_{\rm eff,BO}=+\frac{\Delta(R)}{N}$ has a minimum around $R \sim a_{\bar{3}}$.
We discuss these possibilities in detail in the next subsection. So far, the discussion only involved the two ground states of the LO Hamiltonian, however, similar BO potentials arise for excited states, some of which will be discussed in section \ref{subsec:spinstatistics}.

\subsection{Two types of tetraquarks}\label{sec:twotetraquarks}
Having found the BO potentials, we now discuss the possibility of having 4-quark bound states. We will first show  that for $M_2 \gg N m_{\bar{3}}$ two distinct sets of tetraquarks exist, with states below the two-meson thresholds in both sets. We will then argue that there are no tetraquark bound states, identifiable under the lamppost of the BO approximation, for $M_2 \ll N m_{\bar{3}}$.

 With the BO potential found in the previous subsection and including  the kinetic terms of the heavy quarks, we can now solve for the dependence of the wavefunction of the energy eigenstates on the heavy quark coordinates. This is a two-body problem with a potential dependent only on the relative distance, which can be reduced to a one-body problem with a central potential for the relative coordinate $R$.

\begin{figure}[t]
	\centering 
	\includegraphics[height=3.5cm]{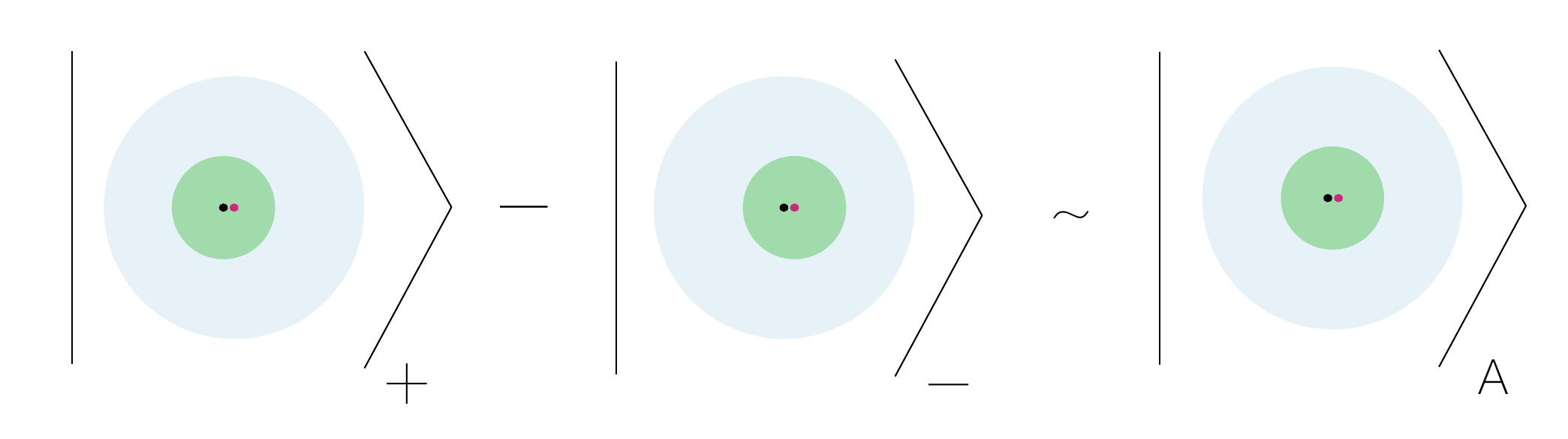}
	\caption{Sketch of type I tetraquark. The black and red dots represent the heavier quarks and the colored regions are where the wavefunction of the lighter antiquarks have a significant support. The ``+" and ``-" subscripts denote the color configuration of the state, as defined in eq. \eqref{plusminusbasis}. We also indicate the large overlap with the state with the two heavy quarks in the antisymmetric color representation.}
	\label{fig::sketch_of_type_I}
\end{figure}

 \textbf{Type-I tetraquarks.} We first consider the $(-)$ branch where the potential $V^-_{\rm eff,BO}=-\frac{\Delta(R)}{N}$ is  attractive at $R \lesssim a_{\bar{3}}$. That can give rise to bound states where the heavy quarks are localised much closer to each other than to the lighter antiquarks.  We call such   states type-I tetraquarks. 
At  $R \ll a_{\bar{3}}$, the BO potential is $\sim -\frac{1}{N}\frac{\alpha}{R}$ so that the possible energy eigenstates would be localised within a radius
 \beq
A_{12} \equiv N (\alpha M_2)^{-1}= a_{\bar{3}} \frac{N m_{\bar{3}}}{M_2}.
\eeq
The BO condition in eq.~(\ref{eq:conditionBO}) reads $A_{12}\ll a_{\bar{3}}$, which  by the above equation  implies $\frac{M_2}{m_{\bar{3}}} \gg N$. The same condition also ensures that the resulting  $\sim \frac{\alpha^2}{N^2}M_2$ binding energy is much larger than  ${\cal E}_{\bar 3}/N$. Consequently the energy of these states
 \beq \label{eq:type_I_energy}
 E_\text{type-I} = E_0 - \mathcal{O}\left(\frac{\alpha^2}{N^2}M_2\right)
 \eeq
  not only falls well below the two meson threshold, but also  below the minimum of the BO potential in the $(+)$ branch and hence below the energy of any bound state that may exist in that other branch.
 Therefore for $M_2 \gg N m_{\bar{3}}$, the ground state of the 4-quark system under study is a tetraquark with the heavier quarks bound together at distances much shorter than the size of the mesons involving the lighter quarks.
Notice that $|{\psi_0^+}\rangle$ and $|{\psi_0^-}\rangle$ have different color structure and generically different coordinate dependence, so that generically  $|{\psi_0^+}\rangle- |{\psi_0^-}\rangle$ is an entangled superposition of color and spacial variables. 
However, the type-I bound states are non-generic superpositions where the two heavy quarks are localized is a small region of size $A_{12}\ll a_{\bar{3}}$. That leads  to   factorization of color and position up to $O(A_{12}/a_{\bar{3}})$ corrections, with the two heavy (and two light) quarks  lying in the anti-symmetric color state (see eq.~(\ref{plusminusbasis})).
A schematic sketch of the type-I tetraquark wavefunctions is shown in Fig. \ref{fig::sketch_of_type_I}.

In the regime of still heavier quarks $M_2 \gg N^2 m_{\bar{3}} $, one can establish the existence of 
 type-I states even without using the BO approximation, see e.g.~\cite{Manohar:1992nd,Eichten:2017ffp}. In this regime, one can first solve the dynamics of system of two heavier quarks, where one finds deeply bound diquark states in the antisymmetric representation. As the quark-quark potential is roughly  $-\frac{\alpha}{N r_{12}}$, the resulting binding energy $E_{12}\sim \alpha^2 \frac{M_2}{N^2}$ dominates over all other possible contributions to the energy of the four quark  system, as they are all $\lesssim \mathcal{E}_{\bar{3}}\sim \alpha^2 m_{\bar{3}}$.
 Moreover, and relatedly, the time scale associated to motion in the diquark system, $\sim \left(\alpha^2 \frac{M_2}{N^2}\right)^{-1}$, is much shorter than  the time scale associated to the 
 motion of the lighter anti-quarks. One can therefore integrate out the diquark dynamics first and then solve the effective dynamics of the system composed of the diquark and the two antiquarks.
 Nevertheless we can find this state consistently also within the BO approximation, as the dynamics of the heavy quarks has negligible influence on the light antiquarks, since they are localized in a small region. This is against the common lore according to which the BO approximation corresponds to  integrating out the faster dynamics of the light particles. 
 The unique color-singlet  configuration out of the antisymmetric diquark  and the two antifundamental antiquarks has a binding potential and leads to a bound state of the three constituents \footnote{For $SU(3)$, the antisymmetric representation coincides with  the antifundamental, so that the unique color singlet contraction of the diquark and the two antiquarks is the same as that of a baryon.}. One can solve easily for the wavefunction of such states;
 at leading order in $1/N$, the antiquarks only interact with the diquark and not with each other (see equation~(\ref{eq:Potential_4quarks_SA}))  so that the problem factorizes into  two ``Hydrogen atoms". The  subleading $1/N$ corrections can be treated perturbatively. That is the same situation of a nucleus with large charge $Ze\gg e$ surrounded by just two electrons.
 
\begin{figure}[t]
	\centering 
	\includegraphics[height=3.7cm]{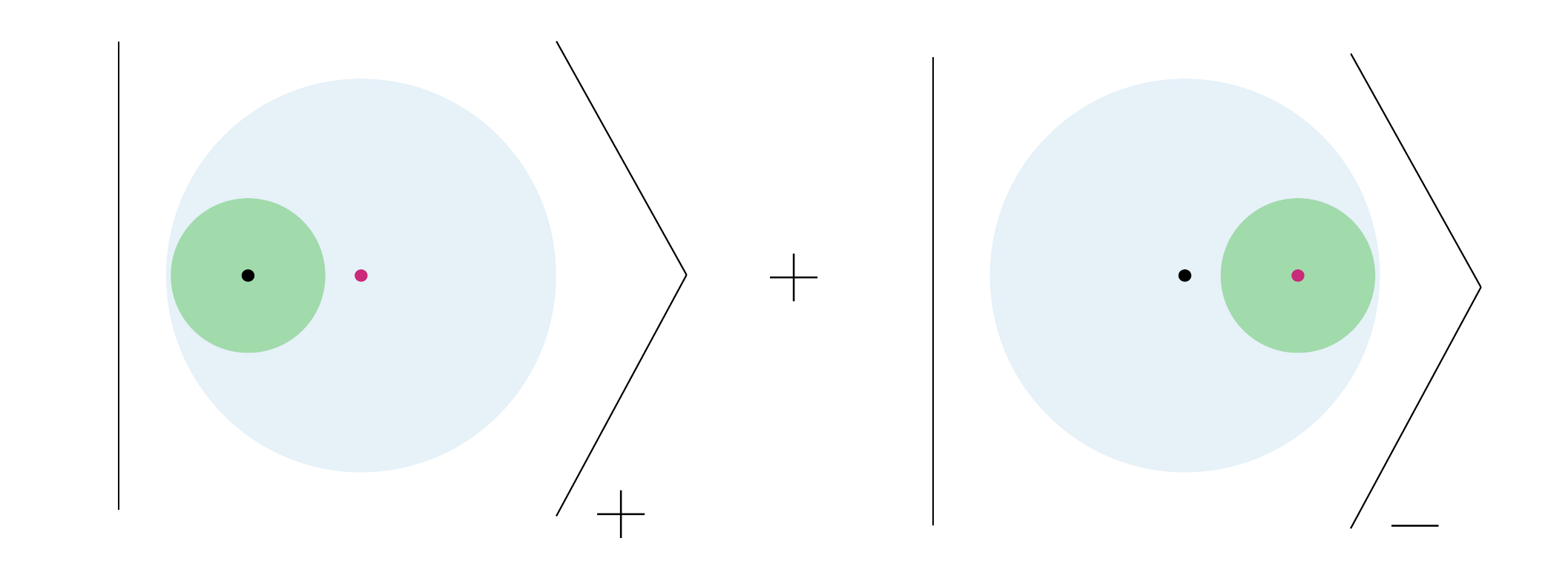}
	\caption{Sketch of the type-II tetraquarks. The black and red dots represent the heavier quarks and the colored regions are where the wavefunction of the lighter antiquarks have a significant support. The ``+" and ``-" subscripts denote the color configuration of the state, as defined in eq. \eqref{plusminusbasis}. This state has no similarity to any state with a fixed color configuration.}
	\label{fig::sketch_of_type_II}
\end{figure}

\textbf{Type-II tetraquarks.} 
Let us consider now the (+) branch where the potential $V^+_{\rm eff,BO}=\frac{\Delta(R)}{N}$ is  repulsive at small $R$ and has a minimum at $R \sim a_{\bar{3}}$.
We refer to the bound states that can possibly arise around the minimum  as type-II tetraquarks.
For such states, the Schr\"odinger equation for the $R$ coordinate is approximately that of a one-dimensional harmonic oscillator, with frequency $\omega \sim \sqrt{\frac{E_0}{N  a_{\bar{3}}^2} \frac{1}{M_2}}\sim \sqrt{\frac{N m_{\bar{3}}}{ M_2}} \frac{E_0}{N}$ set by the BO potential \footnote{For non-zero and $\mathcal{O}(1)$ angular momentum $l$, the contribution of the ``centrifugal" term is small compared to that of the BO potential around its minimum for $M_2 \gg m_{\bar{3}} N $. Hence its effect on the wavefunction for the coordinate $R$ can be neglected. In other words the level separation due to rotational modes is small compared to the vibrational modes of the heavier quarks.}. The low-lying bound states of such a harmonic oscillator are localized within a length $\Delta R\sim \left( \frac{N m_{\bar{3}}}{ M_2}\right)^{1/4} a_{\bar{3}}$ from  the minimum of the potential. The choice $\frac{N m_{\bar{3}}}{ M_2} \ll 1$, then coincidentally implies  the BO condition of eq.~(\ref{eq:conditionBO}) and the validity of the harmonic oscillator approximation for the BO potential around the minimum. That allows to self-consistently identify a set of bound states, with energies given by
\beq
E_{\rm type-II}= E_0 -\mathcal{O}\left( \frac{\mathcal{E}_3}{N}\right)+ \mathcal{O}(\omega),
\eeq 
 below the two-meson thresholds. 
 In figure \ref{fig::sketch_of_type_II}, we show a sketch of the type-II tetraquarks. 
In contrast to
the type-I states, the type-II tetraquarks 
correspond to  a highly entangled superposition in color and coordinate space. In other words they are not in a definite color configuration.

Note that the BO potentials shown in figure \ref{fig:BO_4quarks_eta} are not only suppressed by $1/N$ but have an additional numerical $\mathcal{O}(0.1)$ suppression at the minimum, leading to tetraquarks that are very close to the two meson threshold. 
Remarkably, this numerical suppression can be shown to happen generically and is easily understood given the potential in equation \eqref{eq:potentialequalmass} obtained for $m_{\bar{3}}\gg m_{\bar{4}}$.
To make the discussion 
more clear, consider potentials of similar form
\beq
\label{eq:toypotential}
V_\epsilon(X)= e^{-X} \left( \frac{1}{X} - \epsilon X \right),
\eeq
where we introduced an additional parameter $\epsilon$. The minimum of this potential occurs at $X_{\rm min}\sim 1/\sqrt{\epsilon}$, where it is of order $\sqrt{\epsilon}\, e^{-1/\sqrt{\epsilon}}$. Hence an {\it {algebraically}} small $\epsilon$ leads to an {\it {exponentially}} suppressed energy difference between the tetraquarks and the threshold \footnote{A similar  mechanism 
ensures the exponential suppression of mass scales generated by the slow RG evolution of marginally relevant parameters, like the gauge coupling in QCD or like the Goldberger-Wise dual coupling in the Randall-Sundrum model.}. Even for $\epsilon=2/3$ as in eq. \eqref{eq:potentialequalmass}, the position of the minimum is already at a somewhat large value of $X\simeq 2.07$, leading to a significant suppression from the exponential. Similarly for $m_{\bar{3}}\sim m_{\bar{4}}$, the potential has an overall exponential factor coming from the wavefunction overlap, which again leads to an exponential suppression if the minimum occurs at a large $R/a_{\bar{3}}$.
\\

We have just seen that the condition $M_2\gg N m_{\bar{3}}$ allows to identify two different types  of tetraquarks, for which we could check a posteriori the validity of the BO approximation (see discussion around equation \eqref{eq:BO_condition}). We can now ask more generally if that condition is indeed necessary for the existence of bound states in the BO effective potential. For the case of a particle with mass $M$ in a central potential $V(r)$ such that $V(r)$ is zero at infinity, the following  Bargmann–Schwinger condition~\cite{Bargmann:1952,Schwinger:1961} is \emph{necessary} for the existence of bound states 
\beq \label{eq:BargmannSchwinger}
\int_0^\infty \Theta\left(-V(r)\right) \, r |V(r)| dr \geq \frac{1}{2 M},
\eeq
where the Heaviside theta function is inserted such that the integral is only over the regions where the potential is negative. The condition for the case of  our BO potentials reads parametrically as $M_2 \gtrsim N m_{\bar{3}}$. That means that, without a hierarchy of masses (at least) as large as $N$, there are no four-quark bound states within the BO approximation.

Indeed by studying numerically the Schr\"odinger equation for our BO potentials, we found the critical ratio $\frac{M_{12}}{N m_{\bar{3}}}$ for which ground state tetraquarks are formed, where $M_{12}= \frac{M_1 M_2}{M_1 +M_2}$. In the limit $m_{\bar{3}} \gg m_{\bar{4}}$, the critical ratio is $1.7$ and $0.9$ for respectively  Type-I  and Type-II tetraquarks.  Instead for  $m_{\bar{3}}=m_{\bar{4}}$ we find somewhat larger critical ratios of  $2.4$ and $1.5$ again for respectively  Type-I and Type-II. We note that for parameters around the critical ratios the heavy quarks are not localized in a region $\Delta R \ll a_{\bar{3}}$ and thus cannot be self-consistently  described by  the  BO approximation.  
However, as we will see in section \ref{sec:beyondBO}, for $m_{\bar 3}=O(m_{\bar 4})$, the bound state problem can be easily be studied beyond the domain of validity of the BO approximation. It then turns out that for the specific case $m_{\bar 3}=m_{\bar 4}$,  the BO approximation and the full treatment coincide at leading order in $1/N$ and $m/M$ (see eq.~\eqref{eq:Schrodinger_BBO}). Therefore, in this specific  case, the critical  ratios quoted above can be trusted even though they occur at the edge of validity of the BO approximation.

\textbf{The very special case of very excited states.} The study of tetraquarks  within the  BO approximation so far only considered  light quarks sitting in their  ground state. As discussed in the next section, particle statistics can force some of them to occupy an excited orbital. The condition of applicability of the BO approximation for the lowest excited states is  of course still given by eq.~\eqref{eq:conditionBOmass}. However, one may wonder what happens in the case of very excited orbitals, characterized by principal quantum numbers $n,n'\gg 1$. In order to get an idea we have repeated  the analysis of this section for the case of large $n=n'$. What essentially happens is that the length scale of the BO potential now grows with $n$.  
In particular for type I tetraquarks, the region where the potential behaves like a Coulombic $\propto 1/R$, before having significant overlap suppression extends up $R\sim n^{3/2} a_{\bar{3}}$.
This can be understood as follows: for very small $R$, the overlap is dominated by the last peak of the meson wavefunction located at a distance $\sim n^2 a_{\bar 3}$. This peak has however a width of order $n^{3/2} a_{\bar 3}$ and therefore beyond the distance speficified by the width, the BO potentials drops significantly compared to a $\propto 1/R$ Couloumbic potential. We have confirmed this also numerically. 
At face value this implies bound states exist in a wider range of $M$,
\beq \label{eq:highlyexcitedcondition} M> Nm_{\bar 3}/n^{3/2}.
\eeq
On the other hand, the request of the BO condition eq.~\eqref{eq:conditionBO} implies a slightly tighter  constraint
 \beq M> Nm_{\bar 3}/n,
 \eeq
 which is nonetheless weaker than  eq.~\eqref{eq:conditionBOmass}. These constituent excited states are in reality expected to be unstable, as we do not see any conserved quantum number preventing their decay to either more deply bound tetraquarks, through gluon emission, or to unbound mesons, possibly without gluon emission.
 So we are not sure how much significance to attribute to this result. Finally, considering the type II sector at large $n$ one does not find any extension to the range of validity of the BO approximation.

\subsection{Tetraquarks with identical quarks: spin-statistics and excited states}\label{subsec:spinstatistics}
In this subsection we consider the cases with at least two identical quarks (or antiquarks). For non-identical quarks, all the states constructed so far are allowed, but in the presence of identical particles only the subset  with the suitable transformation properties under permutations  is allowed. 
A general state is described
by a vector wave function $\Psi_{\alpha_1,...\alpha_4,\beta}(\vec r_{ 1}, \vec r_{2},\vec r_{\bar 3}, \vec r_{\bar 4})$  with $\alpha_i$ labelling the spin of each quark and with $\beta=\pm$ labelling the two possible color singlet contractions associated with $P(\pm)^{ij}_{mn}$ (see eq.~(\ref{wfpmlargeN})). 
The constraints from statistics are expressed in terms of the action of the permutation operators, $P_{12}$ and $P_{\bar{3}\bar{4}}$, for respectively the quantum numbers of $Q_{1,2}$ and $\bar q_{3,4}$. Notice that under the permutation of the color indices of either $Q_{1,2}$ or $\bar q_{3,4}$ the color structures $P(\pm)^{ij}_{mn}$ are mapped into each other.
The action of $P_{12}$ on $\Psi_{\alpha_1,...\alpha_4,\beta}(\vec r_{ 1}, \vec r_{2},\vec r_{\bar 3}, \vec r_{\bar 4})$  then consists in $\alpha_1\leftrightarrow \alpha_2$, $\beta\rightarrow -\beta$,  $\vec r_1\leftrightarrow  \vec r_2$. For $P_{\bar{3}\bar{4}}$ one has instead $\alpha_3\leftrightarrow \alpha_4$, $\beta\rightarrow -\beta$, $\vec r_{\bar 3}\leftrightarrow \vec r_{\bar 4}$. 

Using the coordinate basis introduced at the beginning of sect.~\ref{sec:BO}, in the BO approximation it is convenient to pick a basis  of factorized wave functions of the form
\beq \label{eq:wavefunction}
 \chi^{\rm spin}_{\alpha_1,\alpha_2,\alpha_3,\alpha_4}\Phi_{\vec P_{CM}} (\vec R_{CM}) \, \Phi_{\tilde{n}\tilde{l}\tilde{m}}^{\cal A}(\vec{R}) \, \psi^{\cal A}_\beta(\vec{r}^{\, '}_{\bar{3}},\vec{r}^{\, '}_{\bar{4}}; \vec{R}).
\eeq
Here the first factor $\chi^{\rm spin}$ is a coordinate independent vector in spin space. The second describes the motion of the CM and plays no role in our discussion of bound states.  The  third factor  describes a state of the $Q_1{\text{-}}Q_2$ system with orbital quantum number $\tilde n$ and angular   momentum  numbers $\tilde{l}\,,\tilde{m}$
\footnote{Notice that, as the effective BO hamiltonian in eq.~(\ref{BOhamiltonian}) is rotationally invariant, it's eigenstates can be chosen to have definite angular momentum.}. The last factor is the wavefunction for the color (index $\beta$) and for the coordinates of the lighter antiquarks, which as we have seen can be entangled. The label ``${\cal A}$", which is common to the last two factors,  describes the overall color configuration and the orbital configuration of $\bar q_{3,4}$. Hence it also labels  the resulting BO potential and orbital states of the 
$Q_1{\text{-}}Q_2$ system.

Let us consider first the case of identical   
$\bar{q}_{\bar{3},\bar{4}}$. As it turns out, it will in some case be necessary to consider excited states of the reduced $\bar{q}_{\bar 3}$-$\bar{q}_{\bar 4}$ Hamiltonian. We must thus proceed with slightly more generality than in the previous sections, by identifying the symmetries of the reduced Hamiltonian and by  characterizing the ${\cal A}$ quantum numbers.

To identify the symmetries of the reduced Hamiltonian,
where $\vec R$ is treated as a parameter, we should first identify the symmetries of the full Hamiltonian with the kinetic terms of the heavy quarks neglected. These are  rotations 
($SO(3)$), parity ($\Pi$), the full exchange of the light quark quantum numbers $P_{\bar{3}\bar{4}}$ (for $m_{\bar 3}=m_{\bar 4}$),  but also that of the heavy quark quantum numbers $P_{12}$. The latter is not a symmetry of the full Hamiltonian for $M_1\not = M_2$, but the neglect of the kinetic terms restores it. The symmetry of the reduced Hamiltonian is then the subgroup of the above symmetries under which the position vector $\vec R$ is left invariant.
Choosing coordinates such that $\vec R$ is along the 
$z$ axis we then have that the residual symmetries of the reduced Hamiltonian are:
\begin{itemize}
    \item  $SO(2)$ rotations around the $z$ axis: the states of the light antiquarks can then be labeled by the angular momentum in the $z$ direction, $m_z$, and eigenstates of the leading Hamiltonian with different $m_z$ are not mixed by the subleading terms of the Hamiltonian.
    \item Parity in the $(x,y)$-plane: $A_y:(x,y)\to (x,-y)$. Note that $ A_y L_z=-L_zA_y$, corresponding to  $SO(2)\rtimes A_y=O(2)$.  Then the action of $A_y$ on any state with $m_z\neq 0$ gives a corresponding degenerate state with equal and opposite $m_z$. 
    \item The $Z_2$ transformation $\tilde{\Pi}= P_{12} \Pi$. Indeed, as $\vec R\to -\vec R$ under both $P_{12}$ and $\Pi$, their combined action  clearly leaves $\vec R$ invariant. Indeed, thanks to eqs.~(\ref{r3prime},\ref{r4prime}), $\tilde \Pi$ can also be written by combining the action of parity on the antiquarks $\bar \Pi: \vec{r}^{\, '}_{\bar{3},\bar{4}}\to -\vec{r}^{\, '}_{\bar{3},\bar{4}}$ with the exchange of their color indices. In the $\pm$ color wave-function space that corresponds to
\beq \label{eq:definePisymmetry}
\tilde{\Pi}= \begin{pmatrix}
    0 & \bar{\Pi}  \\
    \bar{\Pi} & 0
\end{pmatrix}.
\eeq
Note that $\tilde{\Pi}$ commutes with both $L_z$ and $A_y$. 
 \item $P_{\bar{3}\bar{4}}$.
\end{itemize}

Let us now construct the complete labels $\cal A$ of the light anti-quark states. In the leading $N\to \infty$ approximation the eigenstates of the reduced Hamiltonian, see  eq.~\eqref{eq:Leading_Potential_pm}, are simply pairs of mesons. In an obvious notation, these can be labelled as 
\begin{equation}
    |\pm; \{n,l,m\} , \{n',l',m'\} \rangle \,\equiv |\cal A\rangle,
\end{equation} where, besides the color contraction $\pm$,  the first set of quantum numbers specify the state of $\bar{q}_{\bar{3}}$, and the second set that of $\bar{q}_{\bar{4}}$. For these states, the coordinates $\vec{r}^{\, '}_{\bar{3}}$ and $\vec{r}^{\, '}_{\bar{4}}$ are centered respectively around $-\vec R/2 $ and $\vec R/2$ for the $+$ color configuration (and instead around $\vec R/2 $ and $-\vec R/2$ for the $-$ color configuration).

The action of  $L_z$, $A_y$, $\tilde{\Pi}$ and $P_{\bar{3}\bar{4}}$ in this basis is given by\footnote{The action of $A_y$ corresponds to choosing the standard spherical harmonics for  angular momentum eigenstates,  which satisfy $Y_{l,m}(\theta,-\varphi)=Y_{l,-m}(\theta,\varphi)$.}
\bea
 L_z |\pm; \{n,l,m\} , \{n',l',m'\} \rangle = && (m+m') |\pm; \{n,l,m\} , \{n',l',m'\} \rangle  \\
 A_y |\pm; \{n,l,m\} , \{n',l',m'\} \rangle = && |\pm; \{ n,l,-m \} , \{ n',l',-m' \} \rangle  \\
  \tilde{\Pi} \, |\pm; \{n,l,m\} , \{n',l',m'\} \rangle =&& (-1)^{l+l'} |\mp; \{n,l,m\} , \{n',l',m'\} \rangle \\
  P_{\bar{3}\bar{4}}|\pm; \{n,l,m\} , \{n',l',m'\} \rangle = &&  |\mp; \{n',l',m'\},\{n,l,m\}  \rangle .
\eea 
We have now all the ingredients to construct the states for identical bosonic or fermionic $\bar{q}_{3}$
$\bar{q}_{\bar 4}$. Acting with the projectors
$\frac{1}{2} \left(1 \pm P_{\bar{3}\bar{4}} \right)$ 
we project on states that are symmetric or antisymmetric under exchange the color and the position of $\bar{q}_{\bar 3}$-$\bar{q}_{\bar 4}$
\begin{align}
 {\mathrm{anti-symmetric}} \quad &   \frac{1}{2}\left [|+; \{n,l,m\} , \{n',l',m'\} \rangle -|-; \{n',l',m'\} , \{n,l,m\}   \rangle \right]\label{antisymmetric}\\
 {\mathrm{symmetric}} \quad &   \frac{1}{2}\left [|+; \{n,l,m\} , \{n',l',m'\} \rangle+|-; \{n',l',m'\} , \{n,l,m\}   \rangle \right]. \label{symmetric}
\end{align}
These states have to be combined with a  spin wave-function with the suitable transformation under $P_{\bar{3}\bar{4}}$. Notice that, according to the discussion in the previous sections, in the ground state  $n=n'=1$ and $l=m=l'=m'=0$ the anti-symmetric and symmetric states correspond respectively to type I and type II tetraquarks. Then in the case of identical
fermionic anti-quarks, the total spin of the $\bar{q}_{\bar{3}}-\bar{q}_{\bar{4}}$ system should be  $1$ and $0$ for respectively type-I and type-II tetraquarks.

If instead $\bar{q}_{\bar{3},\bar{4}}$ are identical scalars, the absence of a spin factor leaves as the only option for their ground state
the symmetric combination in eq.~(\ref{symmetric}), corresponding to the type-II tetraquark $|{\psi_0^+}\rangle +|{\psi_0^-}\rangle$. However when considering excited states with $(n,l,m)\not =(n',l',m')$, also type-I tetraquarks are allowed by the statistics. In some situation these may   even constitute the ground state of the full system, as can be seen even bypassing all the careful classification we have been making.
Consider indeed the regime $M_2 \gg N^2 m_{\bar{3}}$ where one can first solve for the $Q_1-Q_2$ diquark bound state  in the antisymmetric color configuration. The resulting binding energy 
dominates over all other contributions, in particular over the binding energies of the lighter antiquarks. One can then construct bound states of the diquark and of the two antiquarks which  are  symmetric under exchange of $\bar{q}_{\bar{3}}$ and $\bar{q}_{\bar{4}}$ and whose energy is obviously lower than that  of the type-II state. Therefore for $M_2 \gg N^2 m_{\bar{3}}$, the ground state is in this class.

\begin{figure}[t!]
    \centering
\includegraphics[width=\linewidth]{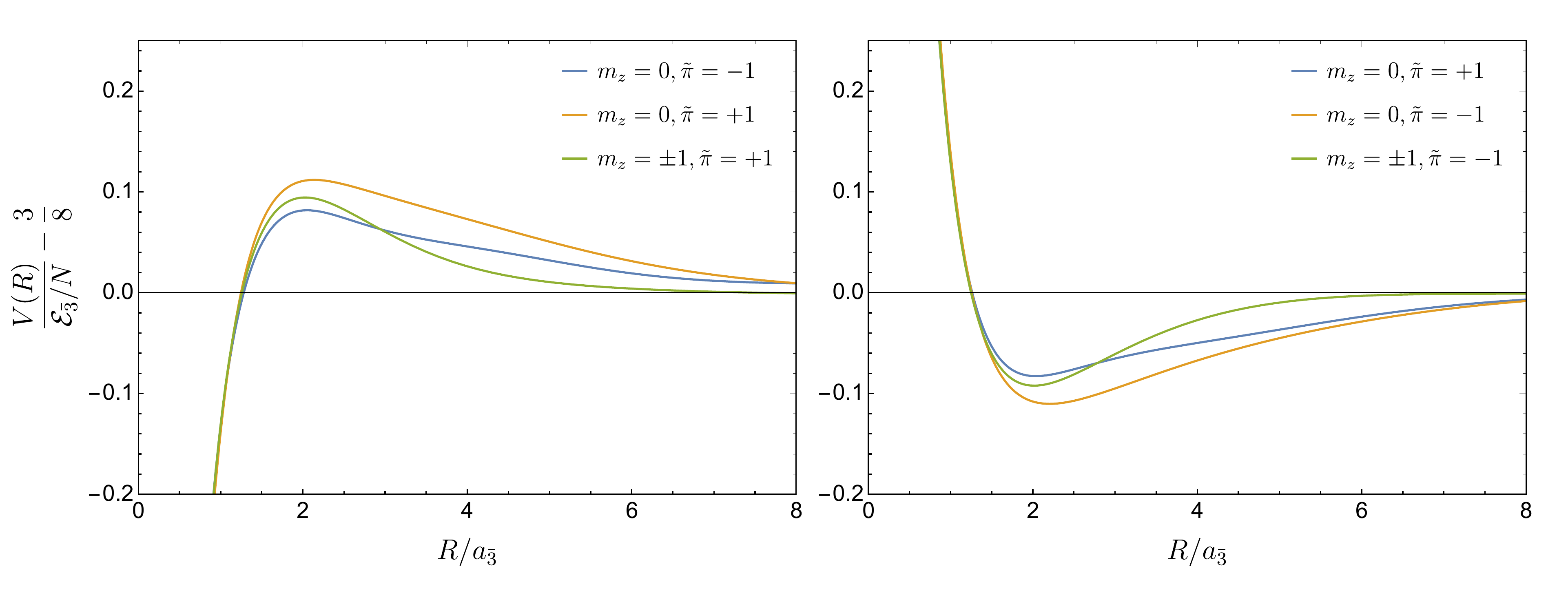}
    \caption{The first excited BO-potentials  for identical antiquarks and $P_{\bar{3}\bar{4}}$-symmetric wavefunctions.  Left: BO potentials which lead to type-I tetraquarks. Right: BO potentials  
    leading to to type-II tetraquarks.
    $\tilde{\pi}$ denotes the eigenvalue of the $\tilde{\Pi}$ transformation.  
    Note that the $m_z=\pm1$ potentials are degenerate on both sides.
    \label{fig:BO_4quarksbosons_excited} }
\end{figure}

To get an idea of the states that arise when considering excited antiquark orbitals, consider the simplest such case $\{n,n' \}=\{ 1,2 \}$.
The  symmetric subset of these states in eq.~(\ref{symmetric}) has an eight-fold degeneracy at leading order in $1/N$: a factor 4  from the  spin states of the $n=2$ orbital and a factor 2 from color.
Now, the  operators $\frac{1}{2}\left(1 \pm \tilde{\Pi} \right)$ project on 4-dimensional subspaces which do not mix even considering higher orders, given $\tilde{\Pi}$ is a symmetry. Each subspace 
features  one state with $m_z=1$, one with $m_z=-1$, and two states with $m_z=0$. Invariance under $SO(2)$  forbids mixing between  $m_z=1$, or $m_z=-1$, with all the other states. Moreover, as $A_y$ maps these states into each other, their energies (and the coresponding BO effective potential)  are degenerate.
Instead the two states with $m_z=0$ in general mix.  At any fixed $R$ the BO potentials are then found by diagonalizing the $1/N$ perturbation  in this two dimensional subspace.
The result is shown in figure \ref{fig:BO_4quarksbosons_excited}. Note that the corresponding states with opposite $\tilde{\Pi}$ quantum numbers lead to BO potentials with the same magnitude but opposite sign. This stems from the fact that the $1/N$ correction to the Hamiltonian is off-diagonal in the basis of eq. \eqref{potentialplusminus}. The shape of these potential makes it evident that there exist both type I and type II tetraquarks  even for identical bosonic $\bar q_{3,4}$ as soon as their excited orbital states are considered.

For identical heavier quarks $Q_{1,2}$,
both type-I and type-II tetraquarks are allowed as 
the full action of the permutation $P_{12}$ now also depends on the angular momentum quantum number $\tilde{l}$ in eq.~(\ref{eq:wavefunction}). Consider indeed for simplicity the ground state $n=n'=1$ of the antiquark orbital. The action of $P_{12}$ on the states in eqs.~(\ref{antisymmetric},\ref{symmetric}) is just a flip of $+$ and $-$: type I ($|{\psi_0^+}\rangle -|{\psi_0^-}\rangle$) has then $P_{12}=-1$ while  II ($|{\psi_0^+}\rangle +|{\psi_0^-}\rangle$) has $P_{12}=+$ . These should be suitably combined with the action  of $P_{12}$ on the obital part, $(-1)^{\tilde l}$, and on the spin part which for fermionic quarks is $(-1)^{S+1}$. Identical bosonic quarks and identical fermionic quarks in the $S=0$ state then feature the same correlation between tetraquark type and angular momentum $\tilde l$:  type-I  eigenstates have odd $\tilde{l}$ while  type-II  have even $\tilde{l}$.
For fermionic quarks in the $S=1$ spin state one has the reverse: type I have even $\tilde{l}$ while type II have odd $\tilde{l}$.

\section{Beyond Born-Oppenheimer}\label{sec:beyondBO}
We  will here study the $Q_1 Q_2\bar{q}_{\bar{3}} \bar{q}_{\bar{4}}$ ground state  supplementing eq.~(\ref{hierarchy}) with
\begin{equation}\label{BBOcondition}
  N m_{\bar{3}}\gg M_2 \gg m_{\bar{3}},
\end{equation}
for which the Born-Oppenheimer approximation fails.
Deriving an effective description for slow moving ground state meson system and applying variant of the Bargmann-Schwinger condition we will show that the ground state consists of two unbound mesons. Indeed our  argument 
is also easily adapted to show that also for other mass hierarchies
 and other mass ordering the ground state consists of a meson pair. In particular we do so for 
the case of a heavy quark and heavy antiquark $Q_1\bar{Q}_{\bar{3}}q_2\bar{q}_{\bar{4}}$ corresponding to  $M_1> M_{\bar{3}}> m_2 >m_{\bar{4}}$.

\subsection{Two heavy quarks and two lighter anti-quarks}
Recall that in the regime of eq.~\eqref{eq:conditionBOmass} where the BO approximation was applicable, we could ignore the kinetic terms of the heavier quarks when solving for the light anti-quark dynamics up to  subleading  $1/N$ order. The heavy quark dynamics was then solved in a second step. In the regime  
of eq.~\eqref{BBOcondition}, which we are here considering, this is no longer possible: $Q_{1,2}$ are here not heavy enough to permit neglecting their recoil effect on the light anti-quark dynamics at order $1/N$.
We must then swap the order at which we include $1/M$ and $1/N$ effects. For the present  analysis, as it will be soon clarified, it is convenient to work in the singlet-adjoint ($I/\text{Ad}$) basis. The Hamiltonian takes the form
\begin{equation}
    H=\frac{P_1^2}{2M_1}+\frac{P_2^2}{2M_2}+\frac{p^2_{\bar{3}}}{2m_{\bar{3}}}+\frac{p^2_{\bar{4}}}{2m_{\bar{4}}}+V^{(0)}+\frac{1}{N}\,V^{(1)}+\mathcal{O}\bigg(\frac{1}{N^2}\bigg).
    \label{1Nhamiltonian}
\end{equation}
where the leading order potential is
\begin{equation}
    V^{(0)}=-\alpha \begin{pmatrix}
        \frac{1}{r_{1\bar{3}}}+\frac{1}{r_{2\bar{4}}} & 0\\
        0 & \frac{1}{r_{1\bar{4}}}+\frac{1}{r_{2\bar{3}}}\\
    \end{pmatrix},
\end{equation}
and where the $1/N$ correction is purely off-diagonal with matrix elements
\begin{equation}
    V_{I,\text{Ad}}^{(1)}= V_{\text{Ad},I}^{(1)}=\alpha\left(\frac{1}{r_{12}}+\frac{1}{r_{\bar{3}\bar{4}}}-\frac{1}{r_{1\bar{4}}}-\frac{1}{r_{2\bar{3}}}\right).
    \label{offdiagAdjI}
\end{equation}
The full potential in this basis is given in eq.~\eqref{SingletAdjointBasis}.
As we shall better explain below, the benefit of the singlet-adjoint basis is that the off-diagonal  potential manifestly falls off faster than $1/R$ at large meson separation.
The leading order system is exactly solvable and consists of two decoupled sectors each containing two non-interacting mesons (and their excited states). We  call these sectors $A$ and $B$, where $A$ is the sector involving $Q_1\bar q_{\bar{3}}$ and $Q_2\bar q_{\bar{4}}$ mesons while $B$ involves $Q_1\bar q_{\bar{4}}$ and $Q_2\bar q_{\bar{3}}$.
To solve the problem, besides the common CM coordinate $\Vec R_{CM}$, it is convenient to use different coordinate bases in sector A and sector B. In sector A we choose $\Vec{r}_{1\bar{3}}$ and $ \Vec{r}_{2\bar{4}}$, which describe the inner dynamics of the $(1\bar{3})$ and $(2\bar{4})$ mesons, and the relative distance between their centers of mass
\begin{equation}
    \vec{R}_{A}=\frac{M_1 \vec{r}_1+m_{\bar{3}} \vec{r}_{\bar{3}}}{M_1+m_{\bar{3}}} - \frac{M_2 \vec{r}_2+m_{\bar{4}} \vec{r}_{\bar{4}}}{M_2+m_{\bar{4}}},
\end{equation}
In sector B, in full analogy, we choose instead $\Vec{r}_{1\bar{4}}$ and $ \Vec{r}_{2\bar{3}}$, as well as
\begin{equation}
    \vec{R}_{B}=\frac{M_1 \vec{r}_1+m_{\bar{4}} \vec{r}_{\bar{4}}}{M_1+m_{\bar{4}}} - \frac{M_2 \vec{r}_2+m_{\bar{3}} \vec{r}_{\bar{3}}}{M_2+m_{\bar{3}}}.
\end{equation}
We indicate the momenta conjugate to $\Vec{R_A}$ and $\Vec{R_B }$ as respectively $\vec{P}_{A}$ and $\vec{P}_B$. The Hamiltonian then becomes 

\begin{equation}
    H= \frac{P_{CM}^2}{2 (M_1+M_2+m_{\bar{3}}+m_{\bar{4}})} + \begin{pmatrix}
        \frac{P_A^2}{2\mu_A}+ \frac{p_{1\bar{3}}^2}{2\mu_{1\bar{3}}}+ \frac{p_{2\bar{4}}^2}{2\mu_{2\bar{4}}} & 0\\
        0 & \frac{P_B^2}{2\mu_B}+ \frac{p_{1\bar{4}}^2}{2\mu_{1\bar{4}}}+ \frac{p_{2\bar{3}}^2}{2\mu_{2\bar{3}}}\\
    \end{pmatrix} + V,
\end{equation}
where $\mu_{i\bar{j}}=\frac{M_i m_{\bar j}}{M_i+m_{\bar j}}$ are the reduced masses corresponding to relative coordinates $\vec{r}_{i\bar{j}}$ and $\mu_{A,B}$ are given by
\beq
\mu_A=\frac{(M_1+m_{\bar{3}})(M_2+m_{\bar{4}})}{M_1+m_{\bar{3}}+M_2+m_{\bar{4}}}, \qquad \text{and} \qquad\mu_B=\frac{(M_1+m_{\bar{4}})(M_2+m_{\bar{3}})}{M_1+m_{\bar{3}}+M_2+m_{\bar{4}}},
\eeq

\begin{figure}[t!]
    \centering
    \includegraphics[scale=0.6]{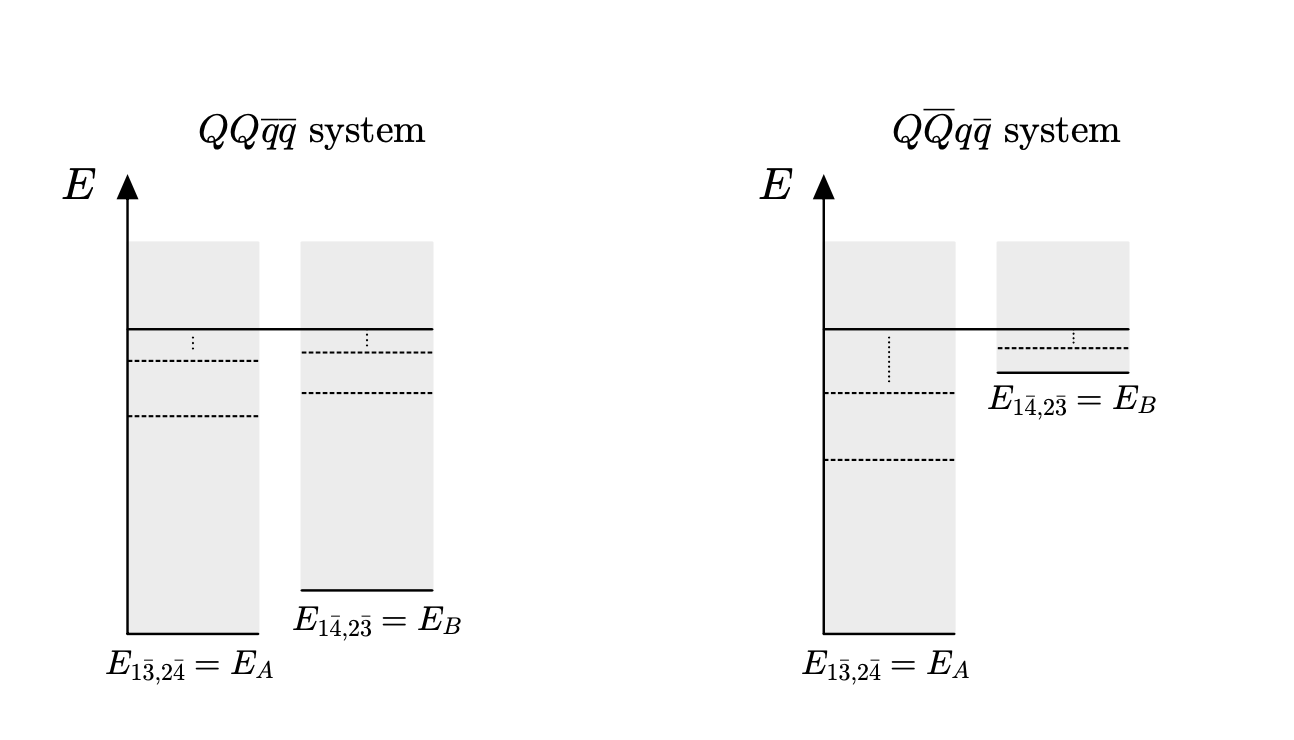}
    \caption{The typical spectrum of the two sectors considering only the interactions at leading order in $1/N$.}
    \label{fig:spectrum}
\end{figure}

According to eq.~(\ref{hierarchy}), the ground states energies in the $A$ and $B$ sector, which we indicate respectively by $E_A$ and $E_B$,  are separated by a positive energy gap 
\begin{equation}
    \Delta E\equiv E_B-E_A\simeq \frac{(M_1-M_2)(m_{\bar 3}-m_{\bar 4})}{M_1M_2}(\mathcal{E}_{\bar 3}+\mathcal{E}_{\bar 4})\,\longrightarrow\,\frac{m_{\bar{3}}}{M_2}\mathcal{E}_{\bar{3}}\,.
    \label{eq:DeltaEgapgeneric}
\end{equation}
In the last step we have taken for illustrative purpose the limit $M_1\gg M_2$, $m_{\bar 3}\gg m_{\bar 4}$, but notice that the gap disappears if either the quarks or the antiquarks are degenerate.
A sketch of the spectrum is shown in the left panel of figure \ref{fig:spectrum}.

The leading order Hamiltonian is diagonal in the basis $\{\ket{A,\vec{P}_A,\alpha_A},\ket{B,\vec{P}_B,\alpha_B}\}$ where the $\alpha_{A,B}$ denote the quantum numbers of the hydrogen-like problem. As such, they can be either discrete, when they describe mesons states, or continuous
\footnote{ Notice that 
 the portion of the excited meson spectrum, discrete or continuous, with absolute value of the energy $\lesssim \Lambda_{QCD}$ is not described by weakly coupled non-relativistic quantum mechanics. Still, as we will now argue, these states decouple from the study of the ground state problem. That is therefore not an issue.}. A general state of the system can be written as
\begin{equation}\label{firstvaransatz}
    \ket{\Psi}=\sum_{\alpha_A}\int d^{3}R_A\,\psi_{\alpha_{A}}(\vec{R}_A)\ket{A,\vec{R}_A,\alpha_A}+\sum_{\alpha_B}
    \int d^{3}R_B\,\psi_{\alpha_{B}}(\vec{R}_B)\ket{B,\vec{R}_B,\alpha_B}.
\end{equation}

The problem is now to study under what condition the $1/N$ correction to the potential in eq.~\eqref{offdiagAdjI} is sufficiently strong a perturbation of the zeroth order Hamiltonian to lead to meson-meson bound states. 
In fact we can ask two different questions: one is whether the lowest energy state is a bound state, the other is whether there exist bound states composed of excited mesons. For this second class of states, as we already mentioned, we do not see any argument for absolute stability, so that we expect them to become metastable when including corrections to the BO approximations and/or when bringing dynamical gluons back into existence. We will
study here  only the first question, though we think our study could easily be extended to that case as well.  We will prove, subject to a  reasonable assumption about the spectrum of excited states, that the ground state of the system is a tetraquark only in  the 
same range where the BO approximation applies,
i.e. for $M_2\gtrsim m_{\bar 3}N$. We think with some more effort we could also prove our additional assumption thus making  our argument complete. However we think that would take us way beyond the  scope of this paper.

In order to proceed it is convenient to first shift the 
 unperturbed $N^0$ Hamiltonian $H_0\to H_0-E_A\equiv H_0'$,  so that the unperturbed ground state has zero energy. Then, as the perturbation $V^{(1)}/N$ vanishes for $R_{A,B}\to \infty$,  the condition for the existence of a stable tetraquark is that the spectrum of the perturbed Hamiltonian  $H'\equiv H_0'+V^{(1)}/N$ extend to negative values. In order to assess that, we should  study the expectation value of $H'$ over the most general class of states in eq.~\eqref{firstvaransatz}. We will not perform this study in full generality but we will work under the reasonable assumption that the lowest energy state of the full Hamiltonian is dominantly a linear superposition of states in the low end of the unperturbed spectrum and study in detail the most general such states\footnote{ Notice that in the limit $N\to \infty$ with all other parameters fixed 
 we expect perturbation theory to apply, in such a way that  eigenvalues above a certain finite  gap will remain positive under the perturbation.}.  Consider now the spectrum in fig. \ref{fig:spectrum}. To simplify the discussion we will work under the assumption that the gap $E_B-E_A$ between the ground states is parametrically smaller than the gap ${\cal E}_{\bar 4}$ to the first excited meson state. For instance we can focus on the case  $m_{\bar 3}\sim m_{\bar 4}$ which implies ${\cal E}_{\bar 3}\sim {\cal E}_{\bar 4}$ and hence $E_B-E_A\ll {\cal E}_{\bar 4}$ according to eqs.~\eqref{BBOcondition} and \eqref{eq:DeltaEgapgeneric} \footnote{ We do not expect our conclusions to be  affected by this hypothesis. For instance, one could  repeat the analysis of this section for the case of just three particles $Q_{1,2}$ and $\bar{q}_{\bar 3}$, which corresponds to the formal  limit $m_{\bar 4}\to 0$ and for which the  unperturbed spectrum consists of states involving one meson and one unbound heavy quark. One would reach the same conclusions. }.
For this choice of parameters, we will then proceed as follows.  We will divide the Hilbert space as the direct sum of two subspaces 
\begin{equation}
\mathcal{H}=\mathcal{H}_{\rm GS}\oplus \mathcal{H}_{\rm exc}\, ,
\end{equation}
where  ${\cal H}_{\rm GS}$ and ${\cal H}_{\rm exc}$ consists respectively of the unperturbed states with energy below and above  the gap ${\cal E}_{\bar 4}$ to the lowest excited meson state.
Labelling  by $|A,\vec{R}_{A},GS\rangle$ and $|B,\vec{R}_{B},GS\rangle$ the ground state meson states for sector A and B respectively, we then have that ${\cal H}_{\rm GS}$ is made up of   the  most general superposition 
\begin{equation}\label{eq:groundstategeneralstate}
    |\Psi\rangle = \int d^3 R_{A}  \,\psi_A(\vec{R}_{A}) |A,\vec{R}_{A},GS\rangle +\int d^3 R_{B} \, \psi_{B}(\vec{R}_{B}) |B,\vec{R}_{B},GS\rangle\,,
\end{equation}
subject to the constraint 
\beq
\frac{P_A^2}{2\mu_A}+\frac{P_B^2}{2\mu_B}<\mathcal{E}_{\bar{4}},
\eeq
corresponding to momenta in the range  
\begin{equation} \label{eq:momentumcutoff}
    P_{_{A,B}}\lesssim \sqrt{M_2 \mathcal{E}_{\bar{4}}} \sim \sqrt{\frac{M_2}{m_{\bar{4}}}} \frac{1}{a_{\bar{4}}}.
\end{equation} 
The complementary subspace  ${\cal H}_{\rm exc}$ consist then, obviousy, of  states  either involving at least one excited mesons or with  kinetic energy exceeding  $ \mathcal{E}_{\bar{4}} $.

The idea is now that bound states, when they first appear as a function of the  parameters, they will approximately consist of linear superposition of states  in  ${\cal H}_{\rm GS}$. To study the problem we can then ``integrate'' out the states in ${\cal H}_{\rm exc}$ and derive an effective Hamiltonian for the reduced ground state Hilbert space ${\cal H}_{\rm GS}$. This procedure is discussed in more detail in appendix \ref{app:effectofexcitedstates}, but the basic implication is easily explained by thinking in terms of standard perturbation theory. As the states in ${\cal H}_{\rm GS}$ have a fixed gap, their contribution to the low energy effective Hamiltonian is quadratic in the perturbation $V^{(1)}/N$ and hence scales like $1/N^2$. This should be compared to the matrix elements of $V^{(1)}/N$ between states in ${\cal H}_{\rm GS}$, which evidently only scale like $1/N$. This different scaling implies (as better detailed in the appendix) that the effects of the virtual excited states is always subdominant for the purpose of assessing the first occurrence of bound states. To study the latter one can then simply study the bound state problem in ${\cal H}_{\rm GS}$ with a Hamiltonian simply given by $H'$ projected to ${\cal H}_{\rm GS}$. The rest of this section is devoted to that.

 We need to compute the matrix elements of
 $V^{(1)}$ on ${\cal H}_{\rm GS}$.
  The potential can be written 
in our Hilbert space basis as $P_{AB} V_{I,Ad}^{(1)}(r_1...r_4)$
where $P_{AB}$ is the operator switching color contraction $A\leftrightarrow B$ and satisfying $P_{AB}^2=\mathbb{I} $.
For the  overlap between the basis states we then find \footnote{Given that $\vec{R}_A = \vec{R}_B + \mathcal{O}\left(\frac{m}{M} \vec{R}_B\right)+ \mathcal{O}\left(\frac{m}{M} \vec{r}_{ij}\right)$, the wave function overlap in eq.~\eqref{ABmatrixelement} is exponentially suppressed unless $|\vec{R}_{A}-\vec{R}_{B}| \lesssim a  m/M$. As this length scale is smaller than than implied by eq.~\eqref{eq:momentumcutoff}, eq.~\eqref{ABmatrixelement} can be well approximated by a $\delta$-function. Notice that our effective theory approach nicely ensures that the perturbation behaves like a potential between point particles.\label{FootnoteDeltaEFT}}
\beq
\langle A,  \vec{R}_{A},GS| P_{AB}|B, \vec{R}_{B},GS\rangle \sim \delta^{3}(\vec{R}_{A}-\vec{R}_{B}) e^{-R_{A}/a_{\bar{3}}} 
e^{-R_{A}/a_{\bar 4}}\, ,
\label{ABmatrixelement}
\eeq
which leads to
\beq\label{Eq:MatrixElementV1}
\langle A,  \vec{R}_{A},GS|V^{(1)}| B, \vec{R}_{B },GS\rangle \simeq \Delta(R_A) \delta^{3}(\vec{R}_{A}-\vec{R}_{B}),
\eeq
up to corrections that are controlled by $m/M$, where $\Delta(R)$ was defined in eq.~\eqref{eq:defineDelta}. On  ${\cal H}_{\rm GS}$ we can then  write the energy functional as
\begin{align}
    \bra{\Psi} H'\ket{\Psi} =&- \int d^3 R_A \psi_A^*(\vec{R}_A) \frac{\nabla^2}{2\mu_A}
    \psi_A(\vec{R}_A) + \int d^3 R_B \psi_B^*(\vec{R}_B) \left( -\frac{\nabla^2}{2\mu_B} +\Delta E\right)
    \psi_B(\vec{R}_B) \nonumber \\
    & \int d^3 R_A \, d^3 R_B \psi_A^*(\vec{R}_A)
    \psi_B(\vec{R}_B) \Delta(R_A) \delta^{3}(\vec{R}_{A}-\vec{R}_{B}) + \text{c.c.},
\end{align}
which leads to the Schr\"odinger equation
\begin{equation} \label{eq:Schrodinger_BBO}
 \left[ -  \begin{pmatrix}
        \frac{\nabla^2}{2\mu_B} & 0\\
        0 & \frac{\nabla^2}{2\mu_B}\\
    \end{pmatrix} +   \begin{pmatrix}
      \frac{\mu_A-\mu_B}{2\mu_A \mu_B } \nabla^2  & 0\\
        0 & \Delta E\\
    \end{pmatrix} + \frac{\Delta(R)}{N} \begin{pmatrix}
       0 & 1\\
      1 & 0 \\
    \end{pmatrix}\right]\begin{pmatrix}
       \psi_A(\vec{R}) \\
       \psi_B(\vec{R}) \\
    \end{pmatrix} = E \begin{pmatrix}
       \psi_A(\vec{R}) \\
       \psi_B(\vec{R}) \\
    \end{pmatrix}.
\end{equation}
The second term in square brackets is a positive semi-definite operator, as $\Delta E \ge 0$ and
\begin{equation}
    \mu_B-\mu_A = \frac{\left(M_1-M_2\right)\left(m_{\bar 3}-m_{\bar 4}\right)}{M_1+M_2+m_{\bar 3}+ m_{\bar 4}}\ge 0.
    \end{equation}
Therefore, it can only increase the ground state energy. We will now study under what condition the modified Hamiltonian that results by dropping this term  has a positive spectrum. A fortiori then, under the same condition also $H'$ is positive definite, at least when reduced to the subspace of eq.~\eqref{eq:groundstategeneralstate}. 

The modified Hamiltonian is particularily simple and can be diagonalized by a basis rotation, $ \psi_{A\pm B}= 1/\sqrt{2} \left(\psi_A \pm  \psi_B \right)$. The system reduces to two decoupled subsectors with   potentials  $\pm \Delta(R)/N$. The application of the Bargmann-Schwinger condition of eq.~\eqref{eq:BargmannSchwinger} to these potential then shows that there are no bound states, i.e. the spectrum is positive, for the mass hierarchy  $N m_{\bar 3} \gg M_2\gg m_{\bar 3}$.
This result is quickly understood. The function $\Delta(R)$ has the form ${\cal E}_{\bar 3}F(R/a_{\bar 3})$
in such a way that the integral at the left hand side  of eq.~\eqref{eq:BargmannSchwinger} is of order ${\cal E}_{\bar 3} a_{\bar 3}^2/N \sim 1/(N m_{\bar 3})$. As the right hand side is $\sim 1/M_2$, our result follows.

\subsection{Alternative quark mass hierarchies and orderings}\label{sec:altmass}

In the mass ordering considered in the previous section, the ground states in the A and B sectors could in principle have been very degenerate. In order to account for the possible compensation of the smallness of the $1/N$ perturbation by the small degeneracy we were then forced to consider an effective low energy description 
(the subspace ${\cal H}_{GS}$) which encompassed both sectors.
We found that for the mass hierarchy of eq.~\eqref{BBOcondition} the ground state is not a tetraquark, essentially because the heavy quarks $Q_1$ and $Q_2$ in this regime are not heavy enough to make   the $O(1/N)$ potential a large perturbation. This result suggest that we should be able to exclude stable tetraquarks also for the case where there is no hierarchy between $M_2$ and $m_{\bar 3}$. Consider indeed the generic case where
$M_1>M_2>M_{\bar 3}>M_{\bar 4}$ with all masses roughly of order $M$. In this case the gap between the A and B ground states  as well as that to first excited meson is $\sim \alpha^2 M$. Proceeding like in the previous section, the study of the effect of the $1/N$ suppressed terms can be carried out by zooming on the effective dynamics on a suitable low energy portion ${\cal H}_{GS}$ of the Hilbert space. The natural choice is to have  
${\cal H}_{GS}$ cover the orbital states of the $A$ sector ground state with energy below the relevant gap $\alpha^2 M$.
The complement ${\cal H}_{exc}$ contains then the whole B sector as well all the excited  mesons plus the continuum in the A sector. The effective potential in the resulting effective description is then of order $1/N^2$ and comes from two sources. The first is the diagonal $1/N^2$ term in the original Hamiltonian. That is easily seen to trivially contribute to just a correction to the binding energy of the $1\bar 3$ and $2\bar 4$
mesons of the A sector, and thus does not influence the existence of stable tetraquarks. The second effect originates from integrating out the B sector, as discussed in the previous section and as detailed in appendix \ref{app:effectofexcitedstates}. One can bound the size of this second contribution under the same reasonable hypothesis we applied previously, i.e. that the $O(\alpha^2 M)$ is not wildly modified by the perturbation. One then finds that the resulting effective potential roughly scales like $\sim (\alpha^2M/N^2)f(R M\alpha)$ with $f$ a fast decreasing function for $R\gg 1/(\alpha M)$ and non-singular at smaller $R$. With this result  the integral on the left hand side of eq.~\eqref{eq:BargmannSchwinger} is roughly $O(1/(N^2 M)$. The criterion for the existence of meson bound states is not passed as soon as $N$ is larger than $O(1)$.

In a similar manner we can investigate the $Q\bar{Q}q\bar{q}$ system corresponding to the mass ordering $M_1> M_{\bar{3}}> M_2> M_{\bar 4}$. That includes in particular the   hierarchical case $M_1\sim M_{\bar 3}=O(M)\gg M_{2}\sim M_{\bar 4}=O(m)$ and the case where all masses are comparable and $O(M)$. 
As the results are the same let us consider for definitness the hierarchical case. 

In the $N\rightarrow \infty$ limit this system again consists of two decoupled sectors of two non interacting meson states.
Again, like in the case we just considered,
there is a large gap between the ground states of the two towers (see the right panel of figure \ref{fig:spectrum}).
In one sector, say $A$, the two mesons correspond to the pairing ($Q\bar{Q}$) and ($q\bar{q}$). The binding energy is dominated by that of the first couple and is of order $\alpha^2 M$. The mesons of the second sector, say $B$, are instead of the form $(Q\bar{q})$ and ($\bar{Q}q$) with a much smaller binding energy of order $\alpha^2 m$. Again we can zoom on a low energy effective description limited to ground state meson states in the A sector,  with kinetic energy below the $\sim \alpha^2 M$ gap. Like previously, the effective potential arises at $O(1/N^2)$  and consists of two contributions. A direct one, which trivially only provides a small correction to the $Q\bar{Q}$ and $q\bar{q}$ meson binding energy, and an indirect one arising from integrating out the $B$ sector. 
Using analogous estimates as in appendix \ref{app:effectofexcitedstates}, we find that
this second contribution scales roughly like
$(\alpha^2 m/N^2)(m/M)^3f(R\alpha m)$, with $f$  smooth at short distances and rapidly decreasing for $R\gg 1/(\alpha m)$.
For such potential the integral on the left hand side of eq.~\eqref{eq:BargmannSchwinger} is roughly $(1/N^2)(m/M)^3 1/m$
which cannot even marginally beat the $1/m$ necessary for the occurrence of a bound state. 
The case of comparable masses is simply obtained  by taking $m\sim M$. The situation here coincides with the case first studied in this section: bound states are not possible as soon as $N$ is bigger than $O(1)$.

We thus conclude that also for these other mass patterns the ground state consists of two mesons.

\section{Discussion and speculations about real-world  tetraquarks} \label{sec:realworld}
Our study focused on the limit where  $N$ is large  and all masses are far above $\Lambda_{\rm QCD}$. Note however, as already pointed out in section \ref{sec:BO},
all our results are basically unaffected even  in the case $m_{\bar{4}}\lesssim\Lambda_\text{QCD}$, as long as 
$m_{\bar{3}}\gg\Lambda_\text{QCD}$ remains satisfied.
Indeed in  that case the effects of $\bar q_{\bar 4}$ in the binding dynamics are negligible, with type I and type II tetraquarks bound by the larger binding energies associated with the three heavier quarks $Q_1, Q_2, {\bar q}_{\bar 3}$. Notice that for this range of masses the excitation spectrum associated with the $\bar q_{\bar 4}$ orbitals, and characterized by energy splittings of order $ \Lambda_\text{QCD}$, is now beyond  perturbative control. The larger splittings associated with the heavier quarks remain however under control. The properties of the states with excited $\bar q_{\bar 3}$ orbitals are the same as discusses in section \ref{sec:BO}. In particular for  $N^2\gg M_2/m_{\bar 3}\gg N$ these states are metastable with respect to  decay into the ground state mesons.

The existence and the properties of type I tetraquarks are also largely unaffected by further taking $m_{\bar 3}$ below $\Lambda_\text{QCD}$, at least for the states where  $Q_1-Q_2$ attraction dominates the binding.
That  is easily seen to correspond to states where $R\ll \Lambda_\text{QCD}^{-1}$, which is realized for $ M_2 \gg N \Lambda_\text{QCD}$. The resulting states belong to the same class of the hadrons with $QQ$ content identified long ago in ref.~\cite{Manohar:1992nd}. On the other hand, for $  \Lambda_\text{QCD}>m_{\bar{3}}\ge m_{\bar{4}}$ the existence of type II tetraquarks depends on the detailed form of the BO potential induced at  distances of order $\Lambda_\text{QCD}^{-1}$ by non-perturbative effects. Whether at small or large $N$ the computation of this quantity can only be done within lattice QCD.

In this section we will try to qualitatively apply the  picture obtained in our study to  real world QCD. In that regard, we 
extrapolate our large $N$ and large mass results to $N=3$ and to the physical masses of $b$ and $c$ quarks. Although there are sizeable corrections, we expect that the qualitative picture is preserved, at least partially.

First we consider the case where all quark masses are
above $\Lambda_\text{QCD}$ with a ``hierarchy" between quarks and antiquark masses.
That is relevant for $bb \bar{c}\bar{c}$ states.
As it was discussed in section \ref{sec:twotetraquarks}, we expect stable tetraquark states to form for $\frac{M}{N m}$ above some critical $\mathcal{O}(1)$ value, which we obtained for various BO potentials.
Incidentally,  in the real world the ratio $\frac{m_b}{N m_c}$ is close to unity, making it hard to draw any robust conclusions. Taking the critical mass ratios obtained, $\frac{m_b}{N m_c}>4.8$ for Type-I and $\frac{m_b}{N m_c}>3$ for Type-II, at face value, we should not expect to see  $bb \bar{c}\bar{c}$ tetraquarks that are stable with respect to strong decays into mesons. Of course we are well aware of the stunt represented by our extrapolation. Notice also, as we discussed at the end of section~\ref{sec:twotetraquarks},   the existence of excited tetraquark sitting above the two meson threshold has a lower critical ration $\frac{M}{N m}$. Of course the size of these excited mesons, given the closeness of $m_c$ becomes quickly of order $\Lambda_\text{QCD}^{-1}$. Also for that reason we have not done a detailed study of the critical ratio for these excited states. Nonetheless by a rough rule of thumb (see eq.~\eqref{eq:highlyexcitedcondition}), we expect the critical ratio $\frac{M}{N m}$ for the existence of the first excited tetraquark to be roughly a factor $2^{-3/2}\sim 1/3$ smaller than  the values quoted above.  That is close to unity, which we take as indication   that metastable $bb \bar{c}\bar{c}$ tetraquarks may well exist.

Next we consider the case where one anti-quark is light, $m_{\bar{4}}< \Lambda_\text{QCD}$. For the case of real world QCD, we may use the results of this regime for the case of  $bb\bar{c}\bar{q}$ states, with $\bar{q}$ being a light anti-quark ($\bar{u}$, $\bar{d}$, or $\bar{s}$). In this case, the BO potential is largely unaffected by  the lightest anti-quark $\bar{q}$, corresponding to the regime $m_{\bar{3}}\gg m_{\bar{4}}$. The critical mass ratio in this case is  lower than the case $m_{\bar{4}}=m_{\bar{3}}$. Again for ground state mesons, we find $\frac{m_b}{N m_c}>3.4$ for type-I and $\frac{m_b}{N m_c}>1.8$ for type-II, making the existence of these states more likely than $bb \bar{c}\bar{c}$. That is even more the case for the tetraquarks with excited $c$-quark orbital, for which the critical ratio is further reduced.

Let us finally  consider the case of two light anti-quarks  below the QCD scale, i.e. the possible $bb \bar{q}\bar{q}$ ($T_{bb}$), $b c\bar{q}\bar{q}$ ($T_{bc}$), and  $c c\bar{q}\bar{q}$ ($T_{cc}$) states. 
In the heavy quark limit, the existence of the type-I states relies only on the short distance region where  the BO potential is dominated by the Coulombic potential among the the two heavy quarks. According to our results stable tetraquarks should then exist for $M_2 \gg N \Lambda_{\rm QCD}$. Taken   at face value for the real world, this result indicates that stable $T_{bb}$ likely  exist, while $T_{bc}$ and $T_{cc}$ lie at the edge. That is qualitatively in agreement with the recent observation of $T_{cc}$ right around the two $DD$ threshold \cite{LHCb:2021vvq}.
Notice also that in the regime $M_2 \gg N \Lambda_{\rm QCD}$ the binding energy of the heavy quarks dominates the $O(\Lambda_\text{QCD})$ contributions from the light quarks, whether in their ground state or whether in an excited state. The  excited tetraquarks then can only cascade decay to the ground state tetraquark through 
gluon  emission (with consequent conversion into light mesons).

$T_{bb}$, $T_{bc}$ and $T_{cc}$, in their type-I
incarnation, correspond to the hadrons with QQ diquark core identified in ref. \cite{Manohar:1992nd} and forming the subject of  the studies in refs. \cite{Eichten:2017ffp, Braaten:2020nwp}. In these papers the masses of this family of tetraquarks was predicted on the basis of HQEFT in conjunction with 
quark-diquark symmetry and using the data for heavy-light mesons and  heavy-light-light baryons.
It is interesting to compare the results of these more systematic studies to the qualitative perspective  we just offered above.
In the case of $T_{cc}$, both analyses find that it lies well above the two meson threshold, in contrast with the experimentally determined value of its mass~\cite{LHCb:2021vvq},
which happens instead to agree with the in principle more rudimentary estimate based on the quark model in ref. \cite{Karliner:2017qjm}. Notice however that both analyses did not account for the finite size of the diquark which, as estimated in \cite{An:2018cln}, could easily give a correction that is comparable to the mismatch with observation.
The same corrections can equally be important for  $T_{bc}$. In the case of $T_{bb}$, however,  not only the predictions \cite{Karliner:2017qjm,Eichten:2017ffp, Braaten:2020nwp} for its mass are significantly below threshold but also finite size effects are reduced, by roughly a $(m_c/m_b)^2\sim 10$ factor. According to these HQEFT + diquark symmetry  a $T_{bb}$ stable under QCD interaction should then definitely exist. In fact, lattice studies, see e.g. \cite{Meinel:2022lzo,Francis:2016hui,Junnarkar:2018twb}, have reached similar conclusions. This all appears in agreement with the more qualitative picture suggested by extrapolating the results of our study.

And what about the possibility for type II $T_{bb}$, $T_{bc}$, $T_{cc}$ tetraquarks? As we already explained,
unlike for type I states, 
their existence is not guaranteed in the realistic case of  light anti-quarks  below the QCD scale. It would hinge instead on the properties of the BO potential which we can only imagine computing through lattice QCD simulations. In fact, the current determination of the potential is not very precise at large separations, and it is unclear if an additional minimum at such separations exists \cite{Meinel:2022lzo}. But if a second minimum were determined to exist that would establish the existence of  type-II tetraquarks  in the $T_{QQ}$ family.
Lacking at the moment such precisely determination, we cannot nonetheless refrain from speculating about this possibility. By accepting it,  we would then have two options, type I and type II, for
the recently discovered $T_{cc}$, as both can accommodate the inferred quantum numbers.  
In the case of future discovery of  $T_{bb}$ (and $T_{bc}$) tetraquarks their types may be distinguished 
by their binding energy. While type-I tetraquarks get more and more bound for heavier constituents, the binding energy for type-II tetraquarks saturates at the minimum of the BO potential. Assuming $T_{cc}$ is a type-I tetraquark, the corresponding $T_{bc}$ would be more bound by  order $ \alpha^2 m_c/N^2\sim \alpha_s^2 m_c $, while the corresponding $T_{bb}$ state would have a binding energy of order $\alpha^2 m_b/N^2\sim \alpha_s^2 m_b$.
On the other hand, in the case of type-II, $T_{bb}$, $T_{bc}$, $T_{cc}$  the binding energies should roughly be the same, as they
becomes mass independent  in the limit of infinite heavy quark mass. Moreover, besides the $1/N$ suppression
of this energy which should survive in the realistic case, in our study we also find an additional accidental  $O(1/10)$, due to the exponential behaviour of the light quark wave function. We have no robust reason for that, but if this accidental suppression were to also survive in the realistic case, then it would significantly help bringing the expected $O(\Lambda_\text{QCD})$ range of the binding energy of $T_{cc}$,  closer to its observed $\sim 0.5$ MeV value.

In our construction of tetraquarks within the BO approximation  we focused on  $QQ{\bar q}{\bar q}$ states. The study of this case is  simpler compared to that of $Q\bar{Q} q{\bar q}$ tetraquarks, even within the $1/N$ expansion. That is because  at leading order in $1/N$ and in the large mass expansion, the two ground states of the reduced Hamiltonian eq.~\eqref{HR} are degenerate. Then, as discussed in section \ref{sec:beyondBO}, the bound state problem can be studied by accounting for  the subleading effects in a truncated low energy Hilbert space around  the ground states.
In the case of a heavy $Q\bar{Q}$ pair,  however, the two different color contractions lead to very different binding energies. As we argued in section \ref{sec:beyondBO} the low energy effective description consists on just one sector, that involving the deeply bound $ Q \bar Q$ meson. It is easy to see that at large $N$ no tetraquarks bound states can form  in this sector.
However, our methodology does not allow us to explore, and thus construct or rule out,  metastable $Q \bar{Q}  q{\bar q}$ tetraquarks. Indeed the problem of finding the BO potentials as a function of the distance between the heavy $Q\bar{Q}$, even though more challenging, is well defined and we leave it for future work. It is interesting to determine whether these potentials admits minima at distances of the order of the size of the $Q\bar{q}$ mesons in which case metastable tetraquarks can form for sufficiently heavy masses of the heavier $Q$ and $\bar{Q}$.
This picture would be in line with the current observed candidate states which are all around the $Q{\bar q}$ meson pair thresholds and can decay to the more bound $(Q\bar{Q},q\bar{q})$ pair of mesons.

\acknowledgments
We would like thank Sergei Dubovsky, Glennys Farrar, Brian Henning, David Kaplan, Matthew Walters,  Marek Karliner, Antonio Polosa, Angelo Esposito, Ivan Polyakov, Abhishek Mohapatra and Marc Wagner for discussions. This work  is partially supported by the Swiss National Science Foundation under contract 200020-213104. RR and SS acknowledge the hospitality of   the Perimeter Institute for Theoretical Physics and  RR  acknowledges support from the Simons Collaboration on Confinement and QCD Strings. ME and SS acknowledge the hospitality of the CERN theory group and SS acknowledges the hospitality of the Center for Cosmology and Particle Physics at NYU.

\newpage

\appendix

\section{Wave functions and the Hamiltonian}\label{app:Wavefunctions}
For the reader's ease, in this appendix, we describe the general structure of the states of a $qq \bar{q}\bar{q}$ system and write down the Hamiltonian in the different bases used in the main text. 

\subsection{The states of a $qq\bar{q}\Bar{q}$ system}
A complete set of quantum numbers of a single (anti-)quark state is given by: the position $x$, the color, the spin, and, when needed, additional internal degrees of freedom such as the flavor. The most general $q_1q_2\bar{q}_{\bar{3}}\bar{q}_{\bar{4}}$ state can thus be written as 
\begin{equation}
    \ket{\Psi}=\sum_{\rho} \int \prod_{k=1}^4 d^3 x_k \, \Psi^{i\,j\,}_{m \,n\,} (x,\rho)\ket{1_i(x_1,\rho_1)\,2_j(x_2,\rho_2)\,\bar{3}^m(x_3,\rho_3)\,\bar{4}^n(x_4,\rho_4)}.
\end{equation}
We collectively denoted the spin and the other internal quantum numbers of the $k$-th particle with the index $\rho_k$. The sum over the color indices is left implicit. The normalization of the ket is chosen so that
\begin{equation}
    \braket{\Phi|\Psi}=\sum_{\rho} \int \prod_{k=1}^4 d^3 x_k \,\Phi_{\,\,\,\,i\,j\,}^{*\,m\,n\,}(x,\rho)\, \Psi^{i\,j\,}_{m \,n\,} (x,\rho).
\end{equation}
As explained in the main text, we are only interested in the two-dimensional subspace of color singlet states. A class of basis can be defined by asking one pair of particles,  either $qq$ or $q\bar{q}$, to sit in a definite color representation $(\mathcal{R})$. Indeed, the second pair must always sit in the conjugate representation to neutralize the color. The wave function can then be expanded as 
\begin{equation}
    \Psi^{i\,j\,}_{m \,n\,} (x,\rho)=\sum_{\mathcal{R}}\Psi_\mathcal{R}(x,\rho) P(\mathcal{R})^{i\,j\,}_{m\,n}
\end{equation}
There are three possible bases of this kind corresponding to three possible pairings: $(12)$, $(1\bar{3})$,$(1\bar{4})$. In the first case, $\mathcal{R}$ can be either the symmetric ($S$) or the anti-symmetric $(A)$ representation while in the others $\mathcal{R}$ is either the singlet $(1)$ or the adjoint (Adj) representation. The normalized color wave functions are then given by
\begin{equation}\label{eq:prep}
\begin{split}
     &P(S)^{i\,j\,}_{m\,n}= \frac{1}{\sqrt{2N(N+1)}}(\delta^i_m\delta^j_n+\delta^i_n\delta^j_m),\hspace{1cm} P(A)^{i\,j\,}_{m\,n}= \frac{1}{\sqrt{2N(N-1)}}(\delta^i_m\delta^j_n-\delta^i_n\delta^j_m),\\
     &P(1_{1\bar{3}})^{i\,j\,}_{m\,n}=\frac{1}{N}\delta^{i}_m\delta^j_n,\hspace{4.2cm}P(\text{Adj}_{1\bar{3}})^{i\,j\,}_{m\,n}=\frac{1}{\sqrt{N^2-1}}\big(\delta^i_n\delta^j_m-\frac{1}{N}\delta^i_m\delta^j_n\big),\\
     &P(1_{1\bar{4}})^{i\,j\,}_{m\,n}=\frac{1}{N}\delta^{i}_n\delta^j_m,\hspace{4.2cm}P(\text{Adj}_{1\bar{4}})^{i\,j\,}_{m\,n}=\frac{1}{\sqrt{N^2-1}}\big(\delta^i_m\delta^j_n-\frac{1}{N}\delta^i_m\delta^j_n\big).
\end{split}
\end{equation}
Each line corresponds to an orthonormalized basis.
Note that the wave function for the adjoint state of the $(1\bar{3})$ ($(1\bar{4})$) pair agrees with the $(1\bar{4})$ ($(1\bar{3})$) singlet to leading order in $N$.
Finally, let us note that the angle between two color states is given by
\begin{equation}
    \cos\theta(\mathcal{R}_1,\mathcal{R}_2)=P(\mathcal{R}_1)^{*\,m\,n\,}_{\,\,\,i\,j}P(\mathcal{R}_2)^{i\,j\,}_{m\,n}
\end{equation}
and can be used to perform the change of basis. 

\subsection{The potential in different bases}
\label{app:potential}
Depending on the regime of the masses of the quarks and antiquarks, the $qq\bar{q}\bar{q}$ system is more easily studied using one particular choice of basis. Here, we collect the different alternatives used in the main text. The general form of the potential is the one in equation (\ref{Energyqqbarx2}). Contracting the color structures with the wave functions previously introduced, we can extract the different matrix elements of the potential in the color singlet subspace. Using a notation where the generators in the full color space are denoted as $T^a_1=T^a\otimes\mathbb{I}\otimes\mathbb{I}\otimes\mathbb{I}$, we have
\begin{equation}
    V_{\mathcal{R}_1,\mathcal{R}_2}=\alpha_s  \sum_{i<j} \frac{\lambda_{ij} (\mathcal{R}_1,\mathcal{R}_2)}{r_{ij}},
\end{equation}
with
\begin{equation}
   \lambda_{ij} (\mathcal{R}_1,\mathcal{R}_2) = P^{*}(\mathcal{R}_1)^{ pq}_{mn}\,\left(T_{(i)}^a T_{(j)}^a\right)_{pq\,kl}^{mn\,rs}\,P(\mathcal{R}_2)^{kl}_{rs},
\end{equation}
where $P(\mathcal{R})$ is given in eq.~\eqref{eq:prep} for the different representations.
\subsubsection*{Symmetric/Anti-Symmetric basis}
When the states are chosen so that the two quarks sit in a definite color representation, either the symmetric or the anti-symmetric, the potential is 

\begin{equation}
\begin{split}
     &V_{SS}=-\alpha_s \frac{(N+2)(N-1)}{4N}\left( \frac{1}{r_{1\bar{3}}} + \frac{1}{r_{2\bar{4}}} + \frac{1}{r_{2\bar{3}}} + \frac{1}{r_{1\bar{4}}} \right) + \alpha_s\frac{N-1}{2N}\left( \frac{1}{r_{12}} + \frac{1}{r_{\bar{3}\bar{4}}} \right),    \\
     &V_{SA}=V_{AS}=-\alpha_s \frac{\sqrt{N^2-1}}{4}  \left( \frac{1}{r_{1\bar{3}}} + \frac{1}{r_{2\bar{4}}} - \frac{1}{r_{2\bar{3}}} - \frac{1}{r_{1\bar{4}}} \right), \\
     &V_{AA}= -\alpha_s \frac{(N-2)(N+1)}{4N} \left( \frac{1}{r_{1\bar{3}}} + \frac{1}{r_{2\bar{4}}} + \frac{1}{r_{2\bar{3}}} + \frac{1}{r_{1\bar{4}}} \right)  -\alpha_s \frac{N+1}{2N}\left( \frac{1}{r_{12}} + \frac{1}{r_{\bar{3}\bar{4}}} \right). 
\end{split}
\end{equation}

\subsubsection*{Singlet/Adjoint basis}
In the basis where the color state of the pair $(1\bar{3})$ is either in the singlet or in the adjoint representation we have the potential 
\begin{equation}\label{SingletAdjointBasis}
\begin{split}
    &V_{II}=-\alpha_s\frac{N^2-1}{2N}\left(\frac{1}{r_{1\bar{3}}}+\frac{1}{r_{2\bar{4}}}\right),\\
    &V_{I\text{Ad}}= V_{\text{Ad}I}= -\alpha_s \frac{\sqrt{N^2-1}}{2N}\left(\frac{1}{r_{1\bar{4}}}+\frac{1}{r_{2\bar{3}}}-\frac{1}{r_{12}}-\frac{1}{r_{\bar{3}\bar{4}}}\right),\\
    &V_{\text{Ad}\text{Ad}}= -\alpha_s\frac{N^2-2}{2N}\left(\frac{1}{r_{1\bar{4}}}+\frac{1}{r_{2\bar{3}}}\right)+\alpha_s\frac{1}{2N}\left( \frac{1}{r_{1\bar{3}}}+\frac{1}{r_{2\bar{4}}}-\frac{2}{r_{12}}-\frac{2}{r_{\bar{3}\bar{4}}}\right).
\end{split} 
\end{equation}
\subsubsection*{$+/-$ basis}
The last convenient basis for studying the system corresponds to a $\pi/4$ rotation of the Symmetric/Anti-Symmetric basis,
\begin{equation}
    \Psi_+=\frac{1}{\sqrt{2}}(\Psi_S+\Psi_A), \hspace{1cm} \Psi_-=\frac{1}{\sqrt{2}}(\Psi_S-\Psi_A).\\
\end{equation}
Differently from the previous ones, neither state corresponds to a definite color configuration for a pair of particles. However, they both approach the singlet in the large $N$ limit.
In this case, the off-diagonal terms of the Hamiltonian are $1/N$ suppressed with respect to the leading diagonal contributions 
\begin{equation}
\begin{split}
    &V_{++}=-\alpha_s \frac{N^2-2+N\sqrt{N^2-1}}{4N}\left(\frac{1}{r_{1\bar{3}}}+\frac{1}{r_{2\bar{4}}}\right)  -\alpha_s \frac{N^2-2-N\sqrt{N^2-1}}{4N}\left(\frac{1}{r_{2\bar{3}}}+\frac{1}{r_{1\bar{4}}}\right)\\&\hspace{1.2cm}-\alpha_s \frac{1}{2N}\left(\frac{1}{r_{12}}+ \frac{1}{r_{\bar{3}\bar{4}}}\right),\\
    &V_{+-}=V_{-+}= \alpha_s \frac{1}{2}\left(\frac{1}{r_{12}}+\frac{1}{r_{\bar{3}\bar{4}}}-\frac{1}{2}\left(\frac{1}{r_{1\bar{3}}}+\frac{1}{r_{1\bar{4}}}+\frac{1}{r_{2\bar{3}}}+\frac{1}{r_{2\bar{4}}} \right)\right),\\
    &V_{--}=-\alpha_s \frac{N^2-2-N\sqrt{N^2-1}}{4N}\left(\frac{1}{r_{1\bar{3}}}+\frac{1}{r_{2\bar{4}}}\right)  -\alpha_s \frac{N^2-2+N\sqrt{N^2-1}}{4N}\left(\frac{1}{r_{2\bar{3}}}+\frac{1}{r_{1\bar{4}}}\right)\\&\hspace{1.2cm}-\alpha_s \frac{1}{2N}\left(\frac{1}{r_{12}}+ \frac{1}{r_{\bar{3}\bar{4}}}\right).\\
\end{split}
\end{equation}

\section{The Born-Oppenheimer approximation}\label{app:BOH2}
In this appendix, we briefly review the Born-Oppenheimer approximation. A general exposition is beyond the scope of this paper and can be found in textbooks (see,  for example \cite{weinberg2015lectures}). We will discuss the main aspects of the method by describing an abelian toy example that shares some of the features of the tetraquark system, namely, a hierarchy of masses and a large charge. While some of the results found in this appendix carry over to the non-abelian case, this is not true for others.

\subsection{A large $N$ analog of Hydrogen molecule ion}\label{app:BOH2explicit}

Consider the system of three electrically charged particles. Two of them have mass $M$ and unit charge while the last one has mass $m$ and charge $-N$.  We work under the assumptions: $M\gg m$ and $N\gg 1$. The particles interact via Coulombic interactions. The Hamiltonian is then
\begin{equation}
    H=\frac{P_1^2}{2M}+\frac{P_2^2}{2M}+\frac{p_3^2}{2m}+\frac{\alpha}{|\vec{R}_1-\vec{R}_2|}-\frac{\alpha N}{|\vec{r}_3-\vec{R}_1|}-\frac{\alpha N}{|\vec{r}_3-\vec{R}_2|},
\end{equation}
where capital letters are used to denote the heavy particle variables. A convenient change of coordinates allows us to decouple the center of mass motion.  The Hamiltonian becomes 
\begin{equation}
    H=\frac{P_{CM}^2}{2(2M+m)}+\frac{P^2}{M}+\frac{p^2}{2\mu}+\frac{\alpha}{R}-\frac{\alpha N}{|\vec{r}+\frac{1}{2}\vec{R}|}-\frac{\alpha N}{|\vec{r}-\frac{1}{2}\vec{R}|},
\end{equation}
with the reduced mass $\mu=2Mm/(2M+m)$.
The separation of scales $M\gg m$ suggests the possibility of integrating out the fast modes $\vec{p}, \vec{r}$ and deriving an effective potential for the slow degrees of freedom described by the variables $\vec{P},{\vec{R}}$.
Let us follow \cite{weinberg_2015} and write the wavefunction of the full system as a superposition of states 
\begin{equation}
    \Phi(\vec{r},\vec{R})=\sum_\alpha \varphi_\alpha(\vec{R}) \psi_\alpha(\vec{r};\vec{R}),
\end{equation}
The functions $\{\psi_\alpha\}$ are the eigenstates of the light particle Hamiltonian that is 
\begin{equation}\label{eq:full_SE}
  \left[\frac{p^2}{2\mu}-\frac{\alpha N}{|\vec{r}+\frac{1}{2}\vec{R}|}-\frac{\alpha N}{|\vec{r}-\frac{1}{2}\vec{R}|}\right]\psi_\alpha(\vec{r};\vec{R})= E_\alpha(\vec{R})\psi_\alpha(\vec{r};\vec{R}).
\end{equation}
They constitute a complete basis for the fast degrees of freedom. The Schr\"odinger equation of the full system is then
\begin{align}
\label{eq:BO_validity_SE}
  \sum_\alpha   \left( \frac{P^2}{M} + \frac{\alpha}{R} + E_\alpha(R)   \right) \varphi_\alpha(\vec{R}) \psi_\alpha(\vec{r};\vec{R}) = E \sum_\alpha \varphi_\alpha(\vec{R}) \psi_\alpha(\vec{r};\vec{R}).
\end{align}
Note that the electronic eigenstates are normalized according to
\begin{equation}
    \int d^3r\,\, \psi_\beta^e(\vec{r};\vec{R})^* \psi_\alpha(\vec{r};\vec{R})=\delta_{\alpha\beta}.
\end{equation}
We multiply eq. \eqref{eq:full_SE} with $\psi_\beta(\vec{r};\vec{R})^*$ and integrate over $r$ to find 
\begin{align}
\label{eq:BO_before_approx}
\begin{split}
 \int d^3r\, \psi_\beta(\vec{r};\vec{R})^* &  \left[2\vec{P} \varphi_\alpha(\vec{R}). \frac{\vec{P}}{M} \psi_\alpha(\vec{r}; \vec{R})+\varphi_\alpha(\vec{R}) \frac{P^2}{M} \psi_\alpha(\vec{r};\vec{R})\right]+ \\
& +\left[\frac{P^2}{M}+V_N(R)+E_\beta(R)\right] \varphi_\beta(R)= E \varphi_\beta(R) .
\end{split}
\end{align}
The Born-Oppenheimer approximation consists in neglecting the terms in the first line with respect to the first term in the second line, i.e. assuming
\begin{equation}
     \int d^3 r  \,\psi_\beta(\vec{r}; \vec{R})^* \left[ 2\vec{P} \varphi_\alpha(\vec{R}) \frac{\vec{P}}{M} \psi_\alpha(\vec{r}; \vec{R})+\varphi_\alpha(\vec{R}) \frac{P^2}{M} \psi_\alpha(\vec{r};\vec{R}) \right] \ll \frac{P^2}{M}\varphi_\alpha(\vec{R}). 
\end{equation}
As we have $P \varphi_\alpha(\vec{R}) \sim P_N \varphi_\alpha(\vec{R})$, where $P_N$ is the typical nucleon momentum, and we generically expect $P\psi (\vec{r}; \vec{R}) \sim p_e \psi(\vec{r}; \vec{R})$ as well as $P^2\psi (\vec{r}; \vec{R}) \sim p_e^2 \psi(\vec{r}; \vec{R})$ with $p_e$ the typical electron momentum\footnote{There are however situations in which $P\psi (\vec{r}; \vec{R}) \ll p_e \psi(\vec{r}; \vec{R})$, however typically one still has $P^2\psi (\vec{r}; \vec{R}) \sim p_e^2 \psi(\vec{r}; \vec{R})$. One such situation is exactly the example described in this appendix.}, this is a good approximation as long as
\begin{equation}\label{eq:BO_condition}
    p_e \ll P_N,
\end{equation}
a condition that we can check a posteriori. 

Thus, in the Born-Oppenheimer approximation, we are left with the reduced nuclear problem with Hamiltonian
\begin{equation}
    H=\frac{P^2}{M}+\frac{\alpha}{R}+V(R),
\end{equation}
with the effective potential computed as the eigenvalue of the  electronic ground state with the nuclei treated as static sources. The large $N$ limit allows to solve for $\psi_\alpha(\vec{r};\vec{R})$, and thus also $V(R)$ perturbatively\footnote{Note that this is the difference with respect to the often discussed $H_2^+$, in which no similar expansion parameter exists. There the electronic system is either solved numerically with the help of cylindrical symmetry, or by making the ansatz of orbitals. In contrast, in the large $N$ limit, we can find the analytic solution in perturbation theory.}.
This comes from the fact that due to the large charge of the light particle, the two heavy particles will self-consistently be localized at distances $R_0$ much shorter than the typical Bohr radius of the light particle $a_0$. Assuming that this is indeed the case, one can easily solve for the wavefunction of the light particle perturbatively and in the end check for self-consistency. To leading order we treat the two heavy particles as being at the same position. The solution for the light particle is then just a Hydrogen wavefunction around a nucleus with charge 2, i.e. with Bohr radius $a_0= 1/(2 N \alpha m)$, and ground state energy 
\begin{equation}
    E_0 = 2 N m \alpha^2
\end{equation} 
This energy is independent of $R$, while the leading $R$ dependence is $\mathcal{O}(R^2)$ and can be found using the first order perturbation theory in terms of the following perturbation Hamiltonian 
\begin{align}
\label{eq:H2+_splitting_H}
     \Delta V \equiv 
                  \frac{2Q\alpha}{r} -  \frac{Q\alpha}{| \vec{r} -\vec{R}/2 |} - \frac{Q\alpha}{| \vec{r} +\vec{R}/2 |},
\end{align}
and one finds
\begin{equation}
    \Delta E(R) = \frac{1}{3}E_0 \frac{R^2}{a_0^2} +  \mathcal{O}\left(E_0 \frac{R^3}{a_0^3}\right) 
\end{equation}
This acts as a BO potential for the heavy particles, and the reduced problem is
\begin{equation}
    \left[ -\frac{\nabla_R^2}{M} + \frac{\alpha}{R}+\Delta E(R)\right] \varphi(\vec{R})=E\varphi(\vec{R}).
\end{equation}
The minimum of the potential is at $R_0 = a_0 \left(\frac{3}{2N}\right)^{1/3}$, thus for  verifying our assumptions of $R_0 \ll a_0$ for $N\gg1$. On top of that, the condition for the validity of the BO approximation can be checked explicitly and is found to be $m/M\ll 1$. This is in contrast to the scaling in the main text, where the BO approximation is only valid for $N m/M \ll 1$. The difference stems from the fact that in the non-abelian case the perturbatively generated potential is down by an additional factor of $N$, while here it is of leading order in the $N$ counting.

\section{Analytic form of the Born-Oppenheimer potential}\label{app:BOpot}

In section \ref{sec:BO} we found an analytic expression for the BO potentials in the limit $\frac{m_{\bar{4}}}{m_{\bar{3}}} \rightarrow 0$. In this appendix, we provide the corrections to this expressions for small but nonzero $\frac{m_{\bar{4}}}{m_{\bar{3}}}$.  
Recall that the BO potentials can be written in terms of $\Delta(R)$ defined in eq.~\eqref{eq:defineDelta} as $V_{\rm BO}= \pm \frac{1}{N} \Delta(R)$.  
The integral expression for $\Delta(R)$ 
is given by
\begin{align}
\begin{split}
    \frac{\Delta(R)}{\mathcal{E}_{\bar{3}}} = \frac{2}{\pi^2 a_{\bar{3}}^5}  \left(\frac{m_{\bar{4}}}{m_{\bar{3}}}\right)^3   \int & d^3 r_3 d^3 r_4 ~ e^{-(r_{1\bar{3}}+r_{2\bar{3}})/a_{\bar{3}}} e^{-\frac{m_{\bar{4}}}{m_{\bar{3}}}(r_{1\bar{4}}+r_{2\bar{4}})/a_{\bar{3}}} \\
    & \left[ \frac{1}{r_{12}} + \frac{1}{r_{\bar{3}\bar{4}}} - \frac{1}{2} \left( \frac{1}{r_{1\bar{3}}} + \frac{1}{r_{2\bar{4}}} + \frac{1}{r_{1\bar{4}}} + \frac{1}{r_{2\bar{3}}} \right)\right ]. 
\end{split}
\end{align}
All the terms have a simple analytic form except the following: 
\begin{align}
    I_{34}(R) = \frac{2}{\pi^2 a_{\bar{3}}^5}  \left(\frac{m_{\bar{4}}}{m_{\bar{3}}}\right)^3   \int  d^3 r_3 d^3 r_4 ~  \frac{1}{r_{\bar{3}\bar{4}}} e^{-(r_{1\bar{3}}+r_{2\bar{3}})/a_{\bar{3}}} e^{-\frac{m_{\bar{4}}}{m_{\bar{3}}}(r_{1\bar{4}}+r_{2\bar{4}})/a_{\bar{3}}}. 
\end{align}
The BO potential can thus be written as
\begin{align}
\label{eq:explicit_pot_appendix}
\begin{split}
     \frac{\Delta(R)}{\mathcal{E}_{\bar{3}}} =  I_{34}(R) + 2e^{-(1+\frac{m_{\bar{4}}}{m_{\bar{3}}})\frac{R}{a_{\bar{3}}}} &\left[ \frac{a_{\bar{3}}}{R} -\frac{1}{3}\left(2+3\frac{m_{\bar{4}}}{m_{\bar{3}}}+3 \left(\frac{m_{\bar{4}}}{m_{\bar{3}}}\right)^2\right)\frac{R}{a_{\bar{3}}} \right.  \\  & \left.-\frac{m_{\bar{4}}}{m_{\bar{3}}}\left(\frac{m_{\bar{4}}}{m_{\bar{3}}}+1\right)\left( \frac{R}{a_{\bar{3}}} \right)^2 -\frac{5}{9} \left(\frac{m_{\bar{4}}}{m_{\bar{3}}}\right)^2\left( \frac{R}{a_{\bar{3}}} \right)^3 \right].
\end{split}     
\end{align}
$I_{34}(R)$ can be computed for  $m_{\bar{4}}/m_{\bar{3}} \ll 1$  in a perturbative expansion in $m_{\bar{4}}/m_{\bar{3}}$. The leading-order term is proportional to $m_{\bar{4}}/m_{\bar{3}}$. This term cancels out exactly with the term of the same order  found in eq. (\ref{eq:explicit_pot_appendix}). The next correction to $I_{34}(R)$ is of order $(m_{\bar{4}}/m_{\bar{3}})^3$. Thus, in the $m_{\bar{4}}/m_{\bar{3}} \ll 1$, the BO potential is found to be
\begin{align}
\begin{split}
     \frac{\Delta(R)}{\mathcal{E}_{\bar{3}}} =  2 e^{-\frac{R}{a_{\bar{3}}}} \left( \frac{a_{\bar{3}}}{R} - \frac{2}{3}\frac{R}{a_{\bar{3}}} - \frac{1}{2}\left(\frac{m_{\bar{4}}}{m_{\bar{3}}}\right)^2\frac{R}{a_{\bar{3}}} + \frac{1}{9}\left(\frac{m_{\bar{4}}}{m_{\bar{3}}}\right)^2 \left( \frac{R}{a_{\bar{3}}} \right)^3   \right) + \mathcal{O}\left(\left(\frac{m_{\bar{4}}}{m_{\bar{3}}}\right)^3\right). 
\end{split}     
\end{align}

\section{Effects of the excited states}\label{app:effectofexcitedstates}
In this appendix we discuss the effects of the excited states by estimating their contribution to the effective Hamiltonian governing the dynamics of the low energy part of the spectrum. In section \ref{sec:beyondBO}, we showed that considering only states constructed as suppositions of the ground state mesons, ground state tetraquarks cannot form for the  hierarchy $M_2 \ll N m_{\bar 3}$.
We will argue in this appendix that including
the contribution of the excited states to the low energy dynamics also does not lead to formation of tetraquark ground states. 
For this argument, in addition to the mass hierarchy $M_2 \ll N m$, we assume that the modification of the spectrum of the excited states due to the potential at subleading $1/N$ orders does not remove their energy gap (from the ground state level of the leading Hamiltonian).   

Using the eigenstates of the Hamiltonian $H^{'}_0$, we can write the  Hilbert space as the direct sum of a ground state sector and a sector of excited state Hilbert space,
$\mathcal{H}=\mathcal{H}_{\rm GS}\oplus \mathcal{H}_{\rm exc}$, and denote the respective projection operators as $\mathbb{P}_{\rm GS}$ and $\mathbb{P}_{\rm exc}$, such that  
\begin{equation} \label{eq:sumofprojectors}
    \mathbb{P}_{\rm GS}+\mathbb{P}_{ES}=\mathbb{I}
\end{equation}
The Hamiltonian can be  represented as    
\begin{equation}
    H^{'}=\begin{pmatrix}
        H_{\rm GS} & H_{\rm mix}\\
         H^{\dag}_{\rm mix} & H_{\rm exc}\\
    \end{pmatrix},
\end{equation}
where $H_{\rm GS}= \mathbb{P}_{\rm GS} H^{'}\mathbb{P}_{\rm GS}$ and $H_{\rm exc}=\mathbb{P}_{\rm exc} H^{'} \mathbb{P}_{\rm exc}$ are the projected Hamiltonian into the two subspaces and the mixing is governed only by the terms of the potential subleading in $1/N$, 
\beq
H_{\rm mix}= \frac{1}{N} \mathbb{P}_{\rm GS} V^{(1)} \mathbb{P}_{\rm exc}+ \mathcal{O}\left( 1/N^2 \right).
\eeq

According to  the standard Green's function approach (see for instance \cite{Marino:2021lne}) time evolution in the the ${\cal H}_{GS}$ low energy subspace is governed by the effective Hamiltonian 
is then given by
\beq \label{eq:Heffgreensfunction}
H_{\rm eff}(E)=H_{\rm GS}+\frac{1}{N^2}\,\mathbb{P}_{\rm GS}\,V^{(1)}\,\mathbb{P}_{\rm exc}\frac{1}{E-H_{\rm exc}+i\varepsilon }\mathbb{P}_{\rm exc}\,V^{(1)}\,\mathbb{P}_{\rm GS}
\eeq
The first term includes the leading Hamiltonian as well as the potential projected in the subspace of ground state, the effects of which were shown to not lead to bound states for  $M_2 \ll N m_{\bar 3}$ in section~\ref{sec:beyondBO}. 
We now focus on the second term which gives the contribution of the  states above the gap to the effective Hamiltonian. 
For simplicity of the presentation, we take $m_{\bar 4}\sim m_{\bar 3}$ and therefore $\mathcal{E}_{\bar 3}\sim \mathcal{E}_{\bar 4}$. The spectrum of the leading order part of  $H_{\rm exc}$, is bounded from below by $E_{\rm exc}\gtrsim \mathcal{E}_{
\bar 3}$. We assume that this gap persists also after the $1/N$ corrections are included. 
With this assumption, the second term above is negative definite and its magnitude can be bounded by
\beq
V^{(1)}\,\mathbb{P}_{ES}\frac{1}{-E+H_{\rm ES}}\mathbb{P}_{ES}\,V^{(1)}\lesssim \frac{V^{(1)} \mathbb{P}_{ES} V^{(1)}}{\mathcal{E}_{\bar 3}}
\eeq
We can estimate the matrix elements of the right hand side of the equation above in the basis $|s, \vec{R}_{s}, GS\rangle$ with $s=A,B$. 
This has vanishing matrix elements between two different sectors $A$ and  $B$  since it has two factors of the potential and the potential $V^{(1)}$ is purely sector-off-diagonal. 
Using eq.~\eqref{eq:sumofprojectors}, it can be split into two terms
\begin{equation} \label{eq:potentialsquaredsplit}
    V^{(1)} \mathbb{P}_{ES} V^{(1)}=V^{(1)} V^{(1)} -V^{(1)} \mathbb{P}_{\rm GS} V^{(1)}
\end{equation}
which we now study separately.
We only quote the expressions for the matrix elements in the A sector for which given the basis we have chosen, with each meson of the A sector in a color-singlet state, the fall-off of the interaction at large distance is manifest. 
Similar results hold for the B sector, although the quick fall off of the interaction at large distance is not manifest in the basis eq.~\eqref{SingletAdjointBasis}

The matrix elements of the first of eq.~\eqref{eq:potentialsquaredsplit} term are  
\begin{equation}
    \langle A, \vec{R}^{'}_A,GS| \left(V^{(1)}\right)^2 | A, \vec{R}_A, GS \rangle \simeq \mathcal{E}^2_{\bar 3}  \, F\left( R_A/a_{\bar{3}} \right)\,\delta^3 (\vec{R}_A-\vec{R}^{'}_A)
\end{equation}
The approximation works in the regime defined by equation eq.~\eqref{eq:momentumcutoff}, see also footnote \ref{FootnoteDeltaEFT}.
The $F(R_A/a_{\bar{3}})$ can be estimated to be
\beq
F\left( R_A/a_{\bar{3}} \right)\sim \begin{cases}
\left(a_{\bar{3}}/R_A\right)^2  & R_A \lesssim a_{\bar{3}}, \\
 \left(a_{\bar{3}}/R_A\right)^6 & R_A \gtrsim a_{\bar{3}}.
\end{cases} 
\eeq
The small $R_A \ll a_3$ is dominated by the contribution of the term proportional to $ 1/r_{12}$  in the potential, while the large $R_A \gg a_3$ can be understood from eq.~\eqref{offdiagAdjI} where for $r_{13}, r_{24} \ll r_{12}$ the  potential has a dipole-dipole interaction $\propto 1/r^3_{12}$. We now find the matrix elements of the second term of eq.~\eqref{eq:potentialsquaredsplit} by inserting a complete basis of states in the ground state mesons sector
\beq
\begin{split}
    \langle A, \vec{R}^{'}_A,GS| &V^{(1)} \mathbb{P}_{\rm GS}V^{(1)}| A, \vec{R}_A, GS \rangle \\ 
&=\int d^3 R_B \langle A, \vec{R}^{'}_A,GS| V^{(1)} | B, \vec{R}_B, GS \rangle \langle B, \vec{R}_B,GS|V^{(1)}| A, \vec{R}_A, GS \rangle \\
&=  \delta^3 (\vec{R}_A-\vec{R}^{'}_A) \, \Delta(R_A)^2
\end{split}
\eeq
where we used eq.~\eqref{Eq:MatrixElementV1}. Note that $\Delta(R_A)^2$ is $1/R_A^2$ for $R_A \ll a_{\bar 3}$ which is dominated by the contribution of the term proportional to $1/r_{12}$ in the potential. This indeed cancels the leading short distance $\propto 1/R_A^2$ contribution of $F(R_A/a_{\bar{3}})$. To see this more clearly, note that the ground state mesons in the basis labelled by $\vec{R}_A$ are approximate eigenstates of the $\propto 1/r_{12}$ term in the potential and hence this term leaves a state in the ground state sector in $\mathcal{H}_{\rm GS}$ so that the action of the projector $\mathbb{P}_{ES}$ gives a vanishing result. For the same reason, there are no terms $\propto 1/R_A$ in the full matrix element. Therefore, the effect of the second term in eq.~\eqref{eq:Heffgreensfunction} is bounded by a potential which in the A sector is estimated as
\beq
\begin{cases}
\frac{\mathcal{E}_{\bar 3}}{N^2}  & R_A \lesssim a_{\bar{3}}, \\
 \frac{\mathcal{E}_{\bar 3}}{N^2}\left(a_{\bar{3}}/R_A\right)^6 & R_A \gtrsim a_{\bar{3}}.
\end{cases}
\eeq
From the Bargmann-Schwinger condition, eq.~\eqref{eq:BargmannSchwinger}, it is then obvious that these contribution cannot lead to formation of the bound states as long as $M_2/m_{\bar 3}\ll N^2$. 
 As already stated above, we only showed the matrix elements in the A sector. For the 
 B sector, the short distance behavior is reproduced identically following the same steps. But the long distance $\propto 1/R^6$ fall off is not manifest since in the basis we have chosen the mesons of the B sector are color-singlets only at leading order in $1/N$. However had we chosen a basis defined by $(Q_1\bar{q}_{\bar{4}})$ and $(Q_2\bar{q}_{\bar{3}})$ pairs being both in the singlet or both in the adjoint color representations, the fall of would be manifest in the effective description for the B sector mesons.

 \bibliographystyle{JHEP}
 \bibliography{biblio.bib}

\end{document}